\documentclass[runningheads]{llncs}
\usepackage{verbatim}
\usepackage[T1]{fontenc}
\usepackage{graphicx}
\usepackage{amssymb}
\usepackage{rotating}
\usepackage{multirow}
\usepackage{color}

\begin{document}
\title{Brain tumor segmentation using synthetic MR images - A comparison of GANs and diffusion models}
%
\titlerunning{Brain tumor segmentation using synthetic MR images}
%
\author{Muhammad Usman Akbar\inst{1,2} \and
Måns Larsson\inst{3} \and Ida Blystad\inst{2,5} \and \\
Anders Eklund\inst{1,2,4}}
%
%
\institute{Department of Biomedical Engineering, Linköping University, Linköping, Sweden  \and Center for Medical Image Science and Visualization (CMIV), Linköping, Sweden \and Eigenvision, Malmö, Sweden \and Division of Statistics \& Machine Learning, Department of Computer and Information Science, Linköping University, Linköping, Sweden \and Department of Radiology in Linköping and Department of Health, Medicine and Caring Sciences, Linköping University, Linköping, Sweden}
%
\maketitle              
\begin{abstract}

Large annotated datasets are required for training deep learning models, but in medical imaging data sharing is often complicated due to ethics, anonymization and data protection legislation. Generative AI models, such as generative adversarial networks (GANs) and diffusion models, can today produce very realistic synthetic images, and can potentially facilitate data sharing. However, in order to share synthetic medical images it must first be demonstrated that they can be used for training different networks with acceptable performance. Here, we therefore comprehensively evaluate four GANs (progressive GAN, StyleGAN 1-3) and a diffusion model for the task of brain tumor segmentation (using two segmentation networks, U-Net and a Swin transformer). Our results show that segmentation networks trained on synthetic images reach Dice scores that are 80\% - 90\% of Dice scores when training with real images, but that memorization of the training images can be a problem for diffusion models if the original dataset is too small. Our conclusion is that sharing synthetic medical images is a viable option to sharing real images, but that further work is required. The trained generative models and the generated synthetic images are shared on AIDA data hub.

\end{abstract}
\section{Introduction}

Medical imaging plays a vital role in the diagnosis and treatment of many diseases, enabling healthcare professionals to understand and visualize the internal structures and functions of the human body. With the advancement of artificial intelligence (AI) the field of medical imaging has seen significant improvements in terms of accuracy, efficiency, and cost-effectiveness. AI techniques such as machine learning and deep learning are commonly applied to medical imaging to, for instance, facilitate early detection and diagnosis of diseases and speedup time consuming segmentations~\cite{litjens2017survey,jiang2017artificial}. For example, radiotherapy treatment planning requires segmentation of the tumor and several organs at risk. It is still common that these segmentations are done manually, and segmentation networks can here be used to reduce the required time for one patient from hours to a few minutes~\cite{wong2020comparing}.  

However, training deep learning models, such as convolutional neural networks (CNNs) and vision transformers, for classification or segmentation normally requires large annotated datasets as the models may have millions or billions of parameters. In computer vision, tremendous progress has been made during the last 10 years, and a crucial resource is the open ImageNet database~\cite{deng2009imagenet} which contains more than 14 million labeled images. Techniques developed in computer vision are rapidly transferred to the medical imaging field, but a major constraint is that access to medical images is much more complicated due to ethics, anonymization and data protection legislation (e.g. the general data
protection regulation (GDPR)). There are several openly available medical imaging datasets, but they are much smaller compared to ImageNet (for example, the human connectome project (HCP) shares 1,100 subjects~\cite{van2013wu}, OpenNeuro shares about 30,000~\cite{markiewicz2021openneuro}, UK
biobank will scan and share 100,000~\cite{littlejohns2020uk}). Furthermore, openly available data are often
anonymized through defacing, can represent selective populations around universities, focus on
healthy controls rather than diseased populations, and are often curated before distribution to eliminate
bad quality data. This limits the potential applicability of any model trained on such data in clinical
settings. Hospitals have records containing immense quantities of medical images, but these records
are often not accessible for research due to regulatory hurdles. 

Generative models, such as generative adversarial networks (GANs) and diffusion models, can today produce very realistic synthetic images, by learning the high dimensional distribution of the training images. A potential solution to facilitate sharing of medical images is therefore to generate and share synthetic images, or more precisely synthetic patients, as GDPR should not apply to medical images which do not belong to a specific person (but further legal research is needed). Recent work has demonstrated that generative models (especially diffusion models) can memorize the training images~\cite{carlini2023extracting,somepalli2023diffusion,akbar2023beware}, meaning that the synthetic images are just copies of the training images. As this questions the validity of sharing synthetic medical images, it is thoroughly discussed at the end of this paper.

To share synthetic medical images, and to motivate further research regarding legal aspects and memorization, it must first be demonstrated that they can be used for training deep learning models with acceptable performance. Due to the growing number of generative image models, one must also select the best model.


\subsection{Related work}

Rankin et al.~\cite{rankin2020reliability} used 19 open health datasets to understand the difference in performance of supervised machine learning models trained on synthetic data compared with those trained on real data, but only used tabular datasets and no image data. Similarly, El Emam et al.~\cite{el2021evaluating} used synthetic tabular data from COVID-19 patients to predict death, and obtained similar performance using synthetic data. Using synthetic images for training CNNs for classification has become popular during recent years~\cite{frid2018gan,guan2019breast,qin2020gan,eilertsen,coyner2022synthetic,azizi2023synthetic}, especially in medical imaging~\cite{yi2019generative} where obtaining large annotated datasets is much more time consuming compared to computer vision. On the other hand, related work on training segmentation networks with synthetic images and corresponding annotations is more limited. To generate synthetic images and the corresponding annotations can be done in at least two ways; jointly as a multi-channel image~\cite{bowles2018gan,pollastri2020augmenting,larsson2022does} or as a two-step process where one model generates a synthetic label (annotation) image and another model generates the medical image from the label image~\cite{guibas2017synthetic,shin2018medical,foroozandeh2020synthesizing,shao2023diffuseexpand}. Bowles et al.~\cite{bowles2018gan} demonstrated that adding synthetic images from a 2D GAN lead to improvements of Dice similarity coefficient between 1 and 5 percent, but did not perform training with only synthetic images. Shin et al.~\cite{shin2018medical} also demonstrated small improvements when adding synthetic images as augmentation. Thambawita et al.~\cite{thambawita2022singan} compared different GANs for generating synthetic colonoscopy images and annotations, but did not use more recent models like StyleGAN or diffusion models. Fernandez et al.~\cite{fernandez2022can} used the two-step approach to generate label images with a diffusion model, and then used SPADE~\cite{park2019semantic} to generate the medical image from the label image. They also applied their models to brain tumor images, but only performed a binary tumor segmentation and did not compare with any other generative models. Furthermore, the generative model was only trained with 1064 slices from 5 subjects.

\subsection{This work}

Here we comprehensively evaluate four 2D GANs (progressive GAN~\cite{karras2018progressive}, StyleGAN 1-3~\cite{karras2019style,karras2020analyzing,karras2021alias}) and a 2D diffusion model~\cite{ho2020denoising,nichol2021improved} for generating brain tumor images and tumor annotations, using two openly available datasets (BraTS 2020 and 2021~\cite{bakas3,bakas4,bakas1,bakas2,menze,baid2021rsna}). We demonstrate that using synthetic images for training segmentation networks (a U-Net~\cite{ronneberger2015u} and a Swin transformer~\cite{liu2021swin}) leads to performance metrics which are slightly lower compared to training with real images, and that sharing synthetic images therefore is a viable option to sharing real images (as long as one verifies that the synthetic images are not too similar to the training images). To the best of our knowledge, no such comprehensive evaluation, requiring more than 2000 GPU days for training all models, has previously been performed for medical image segmentation. The trained generative models and the generated synthetic images are shared on AIDA data hub~\cite{hedlund2020key,dataset}.

\section{Results}

\begin{figure*}[htb]
  \centering
  \includegraphics[width=0.99\textwidth]{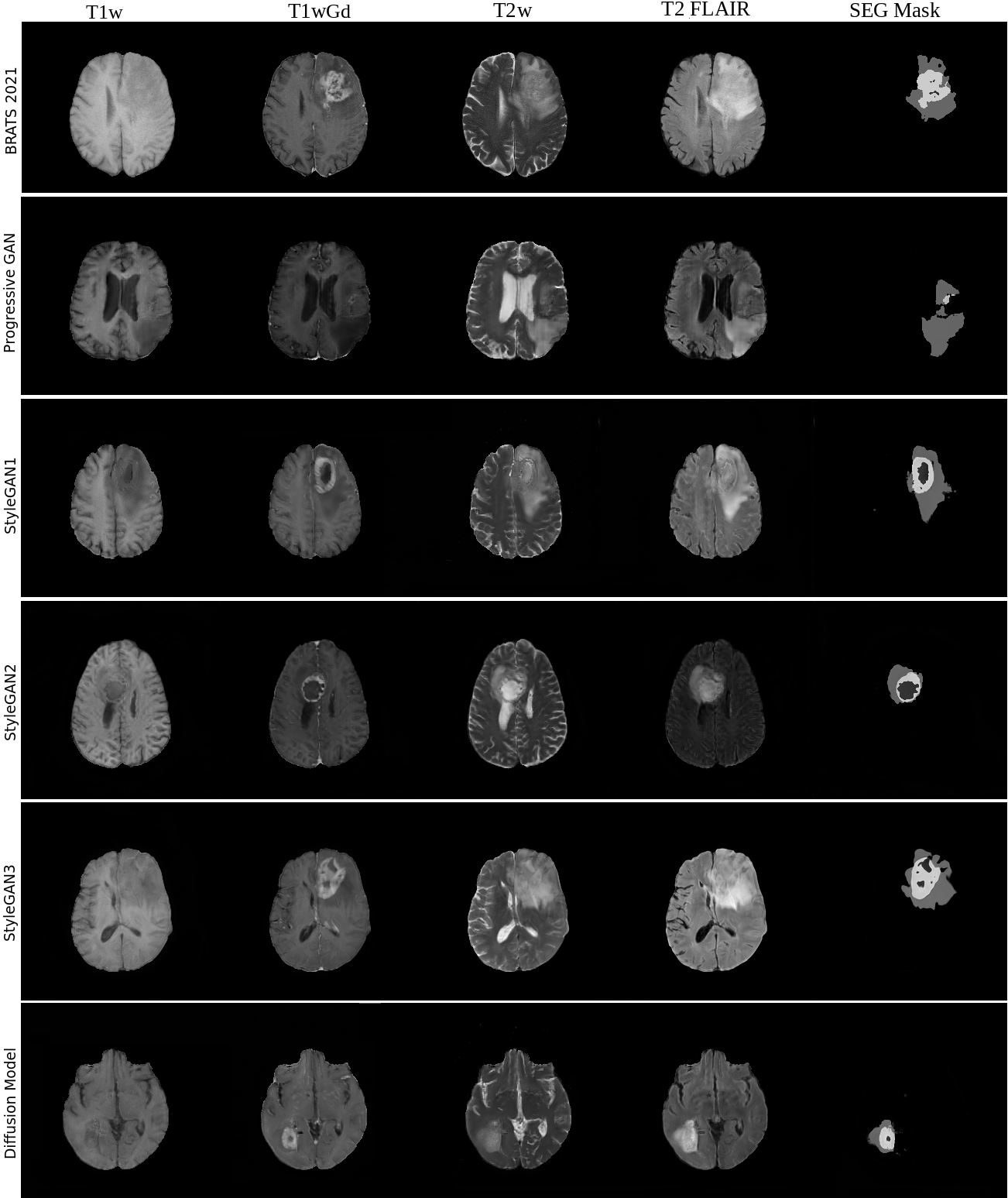} 
  \caption{Synthetic 5-channel images from the BraTS 2021 data. Each row shows a generative model, except for the top row which shows a real example, and each column shows a different MR sequence.}
  \label{fig:Samples_Brats_synthetic_2021}
\end{figure*}

\clearpage

Figure~\ref{fig:Samples_Brats_synthetic_2021} shows a real 5-channel image, and a randomly selected synthetic slice from each generative model. To investigate how distinct the synthetic images are from the training images, we calculated the highest correlation between 100 synthetic images and all training images (see our related work on memorization~\cite{akbar2023beware}). Briefly, synthetic images from a GAN show a distribution of highest correlations which is similar to when comparing training images and test images. For diffusion models, many of the synthetic images are very similar to a training image. Figures~\ref{fig:predictions_2020} -~\ref{fig:predictions_2021} in the Appendix show the resulting U-Net segmentations for a random slice in the two test sets, when training the network with different settings. 

\subsection{Evaluation metrics}

To compare the five generative models we use a variety of metrics. The quality and diversity of synthetic images are often evaluated using metrics such as Frechét inception distance (FID) and inception score (IS)~\cite{borji2019pros}, which use pre-trained CNNs to calculate how different the activations in the CNNs are when feeding real and synthetic images through them. The most important evaluation is in our opinion to train segmentation networks with the synthetic images, and then test how these networks perform on real images. Here we used a U-Net~\cite{ronneberger2015u}, as it is one of the most common networks for medical image segmentation, and a Swin transformer~\cite{liu2021swin}, to see how the results generalize to a more recent network, see the Methods section for details. Segmentation networks are normally evaluated using Dice (measures overlap between true and predicted annotations) and Hausdorff distance (measures the greatest of all the distances from a point in one set to the closest point in the other set). Augmentation is often applied when training segmentation networks, and the segmentation networks were therefore trained with and without augmentation (see Methods section for details). 

\subsection{Ranking of generative models}

To summarize all the results, Table~\ref{tab:ranking} shows the ranking of the five generative models, based on the different metrics FID, IS, Dice and Hausdorff distance. Here we focus on training segmentation networks with synthetic images only, to not make the table too complicated. The diffusion model performs best when comparing the models in terms of Dice and Hausdorff distance, but unfortunately this is in several cases explained by memorization. As expected, the older progressive GAN model often performs worse compared to more recent StyleGAN models. Overall the rankings are similar for U-Net and the Swin transformer. Clearly, the rankings according to the common FID and IS metrics (shown in Table~\ref{tab:models_metrics}) do not correlate well with the ranking according to Dice and Hausdorff distance. Both FID and IS have been questioned as good metrics~\cite{barratt2018note,eilertsen}, but are still commonly used due to the lack of better alternatives. FID and IS focus on image quality and diversity, but do not consider memorization. Since the CNNs used for calculating FID and IS are trained on ImageNet, which only contains non-medical images, the metrics will also be biased for medical images.

\begin{table}\scriptsize 
\begin{tabular}{c|c|c|cccc|}
                                & \textbf{FID}      & \textbf{IS}       & \multicolumn{1}{c|}{\textbf{Dice Aug}} & \multicolumn{1}{c|}{\textbf{Dice}}   & \multicolumn{1}{c|}{\textbf{Hausdorff Aug}} & \textbf{Hausdorff} \\ \hline
\multirow{12}{*}{\rotatebox{90}{\textbf{BraTS 2020}}} & \multirow{4}{*}{} & \multirow{4}{*}{} & \multicolumn{4}{c|}{\textbf{U-Net}}                                                                                                               \\
                                &                   &                   & \multicolumn{1}{c|}{Diffusion}         & \multicolumn{1}{c|}{Diffusion}       & \multicolumn{1}{c|}{Diffusion}              & Diffusion          \\
                                &                   &                   & \multicolumn{1}{c|}{StyleGAN 2}        & \multicolumn{1}{c|}{StyleGAN 3}      & \multicolumn{1}{c|}{Progressive GAN}        & StyleGAN 3         \\
                                &                   &                   & \multicolumn{1}{c|}{StyleGAN 3}        & \multicolumn{1}{c|}{StyleGAN 2}      & \multicolumn{1}{c|}{StyleGAN 2}             & Progressive GAN    \\
                                & Diffusion         & StyleGAN 3        & \multicolumn{1}{c|}{Progressive GAN}   & \multicolumn{1}{c|}{Progressive GAN} & \multicolumn{1}{c|}{StyleGAN 3}             & StyleGAN 2         \\
                                & Progressive GAN   & Progressive GAN   & \multicolumn{1}{c|}{StyleGAN 1}         & \multicolumn{1}{c|}{StyleGAN 1}       & \multicolumn{1}{c|}{StyleGAN 1}             & StyleGAN 1          \\
                                & StyleGAN 1        & Diffusion         & \multicolumn{4}{c|}{\textbf{Swin Transformer}}                                                                                                   \\
                                & StyleGAN 3        & StyleGAN 2        & \multicolumn{1}{c|}{Diffusion}         & \multicolumn{1}{c|}{Diffusion}       & \multicolumn{1}{c|}{Diffusion}              & Diffusion          \\
                                & StyleGAN 2        & StyleGAN 1        & \multicolumn{1}{c|}{StyleGAN 2}        & \multicolumn{1}{c|}{StyleGAN 2}      & \multicolumn{1}{c|}{StyleGAN 2}             & StyleGAN 3         \\
                                & \multirow{3}{*}{} & \multirow{3}{*}{} & \multicolumn{1}{c|}{Progressive GAN}   & \multicolumn{1}{c|}{StyleGAN 3}      & \multicolumn{1}{c|}{StyleGAN 3}             & Progressive GAN    \\
                                &                   &                   & \multicolumn{1}{c|}{StyleGAN 3}        & \multicolumn{1}{c|}{Progressive GAN} & \multicolumn{1}{c|}{Progressive GAN}        & StyleGAN 2         \\
                                &                   &                   & \multicolumn{1}{c|}{StyleGAN 1}        & \multicolumn{1}{c|}{StyleGAN 1}      & \multicolumn{1}{c|}{StyleGAN 1}             & StyleGAN 1         \\ \hline
\multirow{12}{*}{\rotatebox{90}{\textbf{BraTS 2021}}} & \multirow{4}{*}{} & \multirow{4}{*}{} & \multicolumn{4}{c|}{\textbf{U-Net}}                                                                                                               \\
                                &                   &                   & \multicolumn{1}{c|}{Diffusion}         & \multicolumn{1}{c|}{Diffusion}       & \multicolumn{1}{c|}{Diffusion}              & Diffusion          \\
                                &                   &                   & \multicolumn{1}{c|}{StyleGAN 2}        & \multicolumn{1}{c|}{StyleGAN 3}      & \multicolumn{1}{c|}{Progressive GAN}        & StyleGAN 3         \\
                                &                   &                   & \multicolumn{1}{c|}{StyleGAN 3}        & \multicolumn{1}{c|}{StyleGAN 2}      & \multicolumn{1}{c|}{StyleGAN 3}             & Progressive GAN    \\
                                & StyleGAN 2        & StyleGAN 2        & \multicolumn{1}{c|}{StyleGAN 1}        & \multicolumn{1}{c|}{StyleGAN 1}       & \multicolumn{1}{c|}{StyleGAN 1}             & StyleGAN 1          \\
                                & Progressive GAN   & StyleGAN 3        & \multicolumn{1}{c|}{Progressive GAN}   & \multicolumn{1}{c|}{Progressive GAN} & \multicolumn{1}{c|}{StyleGAN 2}             & StyleGAN 2         \\
                                & StyleGAN 1        & Diffusion         & \multicolumn{4}{c|}{\textbf{Swin Transformer}}                                                                                                   \\
                                & Diffusion         & Progressive GAN   & \multicolumn{1}{c|}{Diffusion}         & \multicolumn{1}{c|}{Diffusion}       & \multicolumn{1}{c|}{Diffusion}              & Diffusion          \\
                                & StyleGAN 3        & StyleGAN 1        & \multicolumn{1}{c|}{StyleGAN 2}        & \multicolumn{1}{c|}{StyleGAN 2}      & \multicolumn{1}{c|}{StyleGAN 2}             & StyleGAN 3         \\
                                & \multirow{3}{*}{} & \multirow{3}{*}{} & \multicolumn{1}{c|}{StyleGAN 3}        & \multicolumn{1}{c|}{StyleGAN 3}      & \multicolumn{1}{c|}{StyleGAN 3}             & StyleGAN 2         \\
                                &                   &                   & \multicolumn{1}{c|}{Progressive GAN}   & \multicolumn{1}{c|}{Progressive GAN} & \multicolumn{1}{c|}{Progressive GAN}        & StyleGAN 1         \\
                                &                   &                   & \multicolumn{1}{c|}{StyleGAN 1}        & \multicolumn{1}{c|}{StyleGAN 1}      & \multicolumn{1}{c|}{StyleGAN 1}             & Progressive GAN                  
\end{tabular}
    \vspace{0.2cm}
    \caption{Ranking of the five generative models based on the metrics FID, IS, Dice and Hausdorff distance (when using synthetic images only). Top rows: ranking for BraTS 2020. Bottom rows: ranking for BraTS 2021. For Dice and Hausdorff distance the models are ranked both with and without augmentation when training the segmentation network.}
    \label{tab:ranking}
\end{table}






\begin{table*}\scriptsize
\centering
\begin{tabular}{ccccc}
\textbf{}                            & \multicolumn{2}{c}{\textbf{BRATS 2020}}                         & \multicolumn{2}{c}{\textbf{BRATS 2021}}    \\
\multicolumn{1}{c|}{Model}           & \multicolumn{1}{c|}{FID}      & \multicolumn{1}{c|}{IS}      & \multicolumn{1}{c|}{FID}      & IS      \\ \hline
\multicolumn{1}{c|}{Progressive GAN} & \multicolumn{1}{c|}{18.47098} & \multicolumn{1}{c|}{2.29209} & \multicolumn{1}{c|}{24.58502} & 2.12756 \\
\multicolumn{1}{c|}{StyleGAN 1}      & \multicolumn{1}{c|}{44.85624} & \multicolumn{1}{c|}{2.02602} & \multicolumn{1}{c|}{25.30011} & 2.11348 \\
\multicolumn{1}{c|}{StyleGAN 2}      & \multicolumn{1}{c|}{84.77283} & \multicolumn{1}{c|}{2.20246} & \multicolumn{1}{c|}{20.98637} & 2.22750 \\
\multicolumn{1}{c|}{StyleGAN 3}      & \multicolumn{1}{c|}{58.40920} & \multicolumn{1}{c|}{2.29406} & \multicolumn{1}{c|}{37.49911} & 2.19667 \\
\multicolumn{1}{c|}{Diffusion model} & \multicolumn{1}{c|}{15.85417} & \multicolumn{1}{c|}{2.25878} & \multicolumn{1}{c|}{32.92651} & 2.15857
\end{tabular}%

  \vspace{0.2cm}
    \caption{Comparison of the generative models using the most commonly used metrics, Fréchet inception distance (FID) and inception score (IS). A total of 100,000 synthetic T1wGd images were used to calculate each metric. While these metrics can be calculated rather quickly, they do unfortunately not correlate well with the obtained performance when training networks with synthetic images.}
    \label{tab:models_metrics}
\end{table*}

\subsection{Dice scores}

 Tables~\ref{tab:res_ens_brats2020} and~\ref{tab:res_ens_brats2021} show the obtained Dice scores then training the segmentation networks with different combinations of real and synthetic images, and testing with real images, for BraTS 2020 and BraTS 2021 respectively. To make it easier to compare the performance to only using real images, Table~\ref{tab:relative_dice} shows the relative Dice scores, i.e. the obtained mean Dice score when using real and synthetic images, or only synthetic images, divided by the obtained mean Dice score when using only real images (with augmentation). 

For U-Net trained with BraTS 2020 the diffusion model results in the highest Dice scores when using only synthetic images and augmentation, followed by StyleGAN 2, StyleGAN 3 and progressive GAN. A similar ranking is obtained for the Swin transformer. Using synthetic images from StyleGAN 1 results in very low Dice scores, explained by the fact that we were not able to find good hyperparameters. When excluding StyleGAN 1, the mean Dice score when using synthetic images only is very similar for U-Net and the Swin transformer, demonstrating that the synthetic images can be used also for more recent segmentation networks. Excluding StyleGAN 1, the mean Dice score is improved by 16.8\% for the U-Net when adding augmentation to synthetic images only, compared to 4.1\% for the Swin transformer.

For U-Net trained with BraTS 2021, the diffusion model again results in the highest Dice scores when using only synthetic images and augmentation, followed by StyleGAN 2 and StyleGAN 3. The same ranking is obtained for the Swin transformer. Using synthetic images from StyleGAN 1 results in Dice scores that are much higher compared to for BraTS 2020, possibly explained by the fact that the hyperparameters are a better fit for this dataset. The mean Dice score when using synthetic images only is 6.3\% higher for the Swin transformer compared to the U-Net, again demonstrating that the synthetic images can be used also for more recent segmentation networks. The mean Dice score is improved by 15.9\% for the U-Net when adding augmentation to synthetic images only, compared to only 1.8\% for the Swin transformer. 

Regarding relative Dice scores, Table~\ref{tab:relative_dice} shows that that the diffusion model for U-Net trained with synthetic images only from BraTS 2020 results in the same Dice scores as when using real images, while StyleGAN 2 reaches 66\% - 93\% and StyleGAN 3 reaches 81\% - 87\%. For the Swin transformer, synthetic images from the diffusion model result in Dice scores that are 89\% - 92\% compared to training with real images, while StyleGAN 2 reaches 79\% - 84\% and StyleGAN 3 reaches 78\% - 81\%. For U-Net trained with BraTS 2021 the Dice scores obtained when training with only synthetic images are in general lower compared to BraTS 2020, except for StyleGAN 1. The diffusion model reaches 89\% - 91\% relative Dice, while StyleGAN 2 reaches 63\% - 87\% and StyleGAN 3 reaches 79\% - 82\%. For the Swin transformer the relative Dice scores are in general higher compared to BraTS 2020, partly explained by the fact that the Swin transformer results in a lower Dice score than the U-Net when training with only real images (this may be explained by the fact that vision transformers normally need larger datasets to perform well). The diffusion model reaches about 96\% relative Dice, compared to 85\% - 86\% for StyleGAN 2 and 82\% - 84\% for StyleGAN 3.

\begin{figure*}[htb] 
  \centering
  \includegraphics[width=0.9\textwidth]{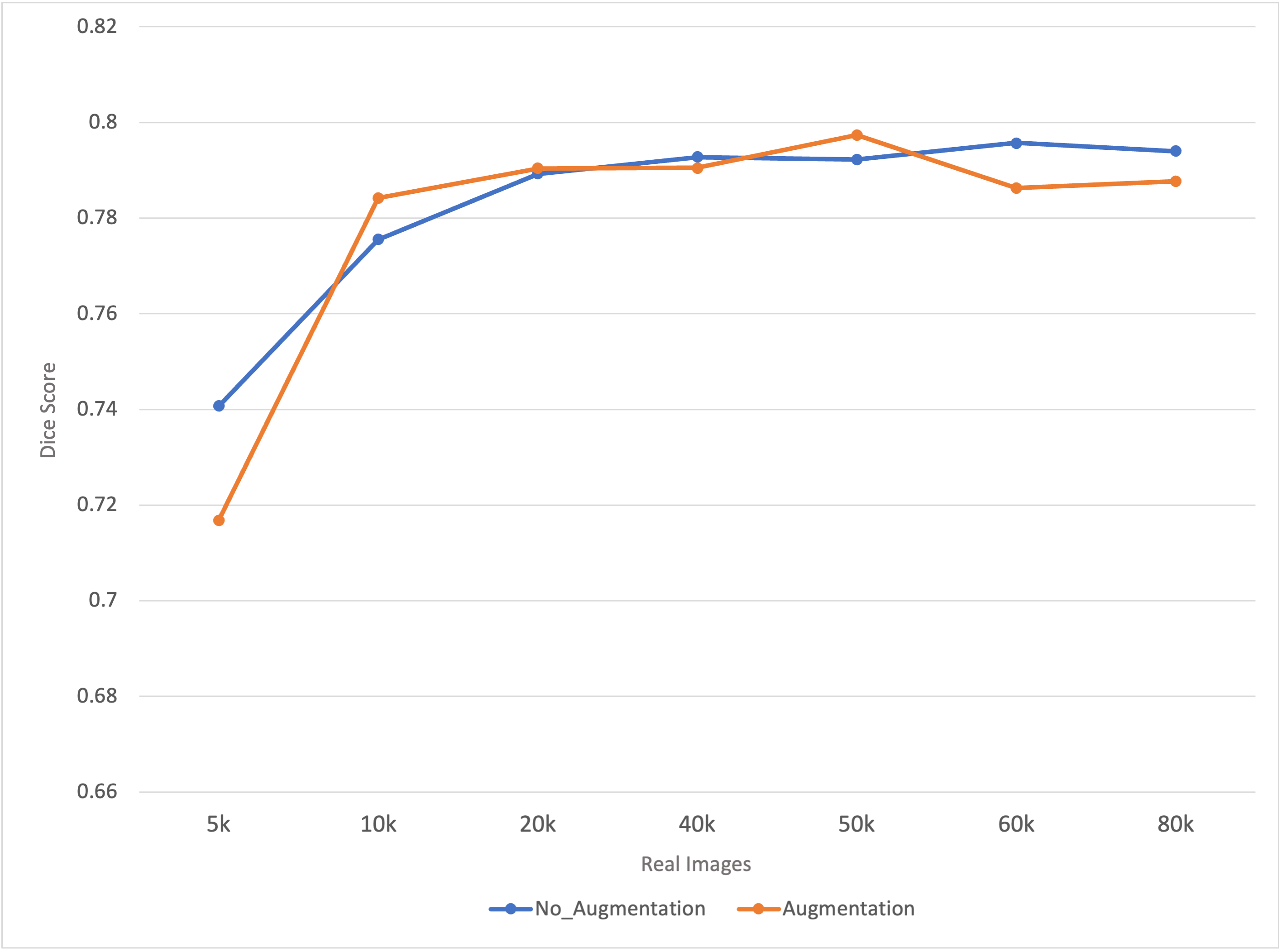} 
  \caption{Graph depicting the U-Net segmentation performance (Dice score) when using different proportions of real (BraTS 2021) and synthetic images generated from StyleGAN 3 (trained on BraTS 2021), in a constant total set of 100,000 images. As the number of real images increases along the x-axis, fewer synthetic images are used. To avoid random fluctuations, each segmentation model was trained 10 times and the average performance is presented.} 
  \label{fig:Orignal_ratio}
\end{figure*}

To assess the impact of the ratio of real and synthetic images, we systematically increased the proportion of real images in a training set with a constant size of 100,000 images. This approach allowed us to evaluate the benefits of real data and the utility of synthetic images in enhancing model performance. The outcomes of this incremental integration are illustrated in Figure~\ref{fig:Orignal_ratio}, which showcases how varying the ratio of real to synthetic data affects the results. Using only 5000 real images, along with 95,000 synthetic images, still results in good performance (substantially higher compared to using 100,000 synthetic images).

\begin{table*}  \scriptsize
    \centering
    \begin{tabular}{c|c|c|c|ccc|c} 
    
\multirow{22}{*}{\rotatebox{90}{\textbf{U-Net}}}
&        \textbf{Model} & \textbf{Orig} & \textbf{Aug} & \textbf{ET} & \textbf{ED} & \textbf{NCR/NET} & \textbf{Mean} \\ \hline
&None                & \checkmark    & \checkmark    & $0.791 \pm 0.009$ & $0.785 \pm 0.003$ & $0.610 \pm 0.008$ & $0.729 \pm 0.004$ \\
&Progressive GAN     & \checkmark    & \checkmark    & $0.790 \pm 0.010$ & $0.790 \pm 0.007$ & $0.609 \pm 0.012$ & $0.730 \pm 0.008$ \\
&StyleGAN 1          & \checkmark    & \checkmark    & $0.769 \pm 0.033$ & $0.759 \pm 0.047$ & $0.593 \pm 0.032$ & $0.707 \pm 0.037$ \\
&\textbf{StyleGAN 2} & \checkmark    & \checkmark    & $\bf{0.802 \pm 0.009}$ & $\bf{0.794 \pm 0.005}$ & $\bf{0.614 \pm 0.009}$ & $\bf{0.737 \pm 0.006}$ \\
&StyleGAN 3          & \checkmark    & \checkmark    & $0.769 \pm 0.042$ & $0.743 \pm 0.060$ & $0.582 \pm 0.062$ & $0.698 \pm 0.054$ \\
&Diffusion           & \checkmark    & \checkmark    & $0.797 \pm 0.020$ & $0.785 \pm 0.025$ & $0.616 \pm 0.014$ & $0.733 \pm 0.019$ \\
&None                & \checkmark    &               & $0.787 \pm 0.010$ & $0.783 \pm 0.004$ & $0.606 \pm 0.009$ & $0.726 \pm 0.005$ \\
&Progressive GAN     & \checkmark    &               & $0.786 \pm 0.010$ & $0.784 \pm 0.011$ & $0.608 \pm 0.010$ & $0.726 \pm 0.008$ \\
&StyleGAN 1          & \checkmark    &               & $0.783 \pm 0.028$ & $0.769 \pm 0.035$ & $0.602 \pm 0.025$ & $0.718 \pm 0.029$ \\
&StyleGAN 2          & \checkmark    &               & $0.799 \pm 0.013$ & $0.788 \pm 0.008$ & $0.611 \pm 0.011$ & $0.732 \pm 0.009$ \\
&StyleGAN 3          & \checkmark    &               & $0.777 \pm 0.031$ & $0.764 \pm 0.047$ & $0.599 \pm 0.048$ & $0.713 \pm 0.041$ \\
&Diffusion           & \checkmark    &               & $0.794 \pm 0.017$ & $0.785 \pm 0.017$ & $0.613 \pm 0.011$ & $0.731 \pm 0.014$ \\
&Progressive GAN     &               &  \checkmark   & $0.668 \pm 0.016$ & $0.607 \pm 0.048$ & $0.468 \pm 0.025$ & $0.581 \pm 0.026$ \\
&StyleGAN 1          &               &  \checkmark   & $0.029 \pm 0.017$ & $0.167 \pm 0.107$ & $0.071 \pm 0.036$ & $0.089 \pm 0.048$ \\
&StyleGAN 2          &               &  \checkmark   & $0.750 \pm 0.007$ & $0.732 \pm 0.013$ & $0.553 \pm 0.018$ & $0.678 \pm 0.009$ \\
&StyleGAN 3          &               &  \checkmark   & $0.714 \pm 0.021$ & $0.709 \pm 0.047$ & $0.497 \pm 0.057$ & $0.640 \pm 0.041$ \\
&\textbf{Diffusion}  &               &  \checkmark   & $\bf{0.791 \pm 0.006}$ & $\bf{0.790 \pm 0.005}$ & $\bf{0.610 \pm 0.006}$ & $\bf{0.730 \pm 0.004}$ \\
&Progressive GAN     &               &               & $0.546 \pm 0.129$ & $0.457 \pm 0.159$ & $0.360 \pm 0.113$ & $0.454 \pm 0.130$ \\
&StyleGAN 1          &               &               & $0.022 \pm 0.014$ & $0.129 \pm 0.099$ & $0.057 \pm 0.032$ & $0.069 \pm 0.044$ \\
&StyleGAN 2          &               &               & $0.562 \pm 0.208$ & $0.499 \pm 0.238$ & $0.376 \pm 0.190$ & $0.479 \pm 0.206$ \\
&StyleGAN 3          &               &               & $0.682 \pm 0.039$ & $0.658 \pm 0.067$ & $0.444 \pm 0.073$ & $0.595 \pm 0.059$ \\
&Diffusion           &               &               & $0.784 \pm 0.009$ & $0.783 \pm 0.008$ & $0.604 \pm 0.008$ & $0.723 \pm 0.008$ \\ \hline
\multirow{22}{*}{\rotatebox{90}{\textbf{Swin Transformer}}}
&None                & \checkmark    & \checkmark    & $0.793 \pm 0.015$    & $0.803 \pm 0.004$  & $0.631 \pm 0.005$   & $0.743 \pm 0.008$ \\
&\textbf{Progressive GAN}     & \checkmark    & \checkmark    & $\textbf{0.772} \pm \textbf{0.005}$    & $\textbf{0.776} \pm \textbf{0.002}$  & $\textbf{0.580} \pm \textbf{0.005}$   & $\textbf{0.709} \pm \textbf{0.003}$ \\
&StyleGAN 1          & \checkmark    & \checkmark    & $0.754 \pm 0.005$    & $0.749 \pm 0.003$  & $0.566 \pm 0.007$   & $0.690 \pm 0.005$ \\
&StyleGAN 2          & \checkmark    & \checkmark    & $0.770 \pm 0.002$    & $0.759 \pm 0.002$  & $0.577 \pm 0.006$   & $0.702 \pm 0.006$ \\
&StyleGAN 3          & \checkmark    & \checkmark    & $0.755 \pm 0.008$    & $0.751 \pm 0.002$  & $0.574 \pm 0.007$   & $0.693 \pm 0.006$ \\
&Diffusion           & \checkmark    & \checkmark    & $0.754 \pm 0.009$    & $0.755 \pm 0.002$  & $0.574 \pm 0.003$   & $0.694 \pm 0.005$ \\
&None                & \checkmark    &               & $0.740 \pm 0.009$    & $0.731 \pm 0.003$  & $0.540 \pm 0.005$   & $0.670 \pm 0.006$ \\
&Progressive GAN     & \checkmark    &               & $0.714 \pm 0.011$    & $0.733 \pm 0.003$  & $0.548 \pm 0.005$   & $0.665 \pm 0.006$ \\
&StyleGAN 1          & \checkmark    &               & $0.732 \pm 0.009$    & $0.736 \pm 0.001$  & $0.539 \pm 0.004$   & $0.669 \pm 0.005$ \\
&StyleGAN 2          & \checkmark    &               & $0.742 \pm 0.012$    & $0.742 \pm 0.002$  & $0.547 \pm 0.004$   & $0.677 \pm 0.005$ \\
&StyleGAN 3          & \checkmark    &               & $0.745 \pm 0.008$    & $0.738 \pm 0.002$  & $0.562 \pm 0.006$   & $0.681 \pm 0.005$ \\
&Diffusion           & \checkmark    &               & $0.749 \pm 0.010$    & $0.740 \pm 0.003$  & $0.543 \pm 0.004$   & $0.677 \pm 0.006$ \\
&Progressive GAN     &               & \checkmark    & $0.667 \pm 0.009$    & $0.676 \pm 0.004$  & $0.476 \pm 0.003$   & $0.607 \pm 0.003$ \\
&StyleGAN 1          &               & \checkmark    & $0.010 \pm 0.002$    & $0.149 \pm 0.003$  & $0.037 \pm 0.006$   & $0.065 \pm 0.004$ \\
&StyleGAN 2          &               & \checkmark    & $0.745 \pm 0.005$    & $0.715 \pm 0.003$  & $0.426 \pm 0.003$   & $0.629 \pm 0.004$ \\
&StyleGAN 3          &               & \checkmark    & $0.713 \pm 0.009$    & $0.694 \pm 0.002$  & $0.400 \pm 0.003$   & $0.603 \pm 0.005$ \\
&\textbf{Diffusion}           &               & \checkmark    & $\textbf{0.754} \pm \textbf{0.007}$    & $\textbf{0.743} \pm \textbf{0.003}$  & $\textbf{0.556} \pm \textbf{0.007}$   & $\textbf{0.684} \pm \textbf{0.006}$ \\
&Progressive GAN     &               &               & $0.659 \pm 0.003$    & $0.639 \pm 0.005$  & $0.458 \pm 0.004$   & $0.585 \pm 0.004$ \\
&StyleGAN 1          &               &               & $0.024 \pm 0.004$    & $0.112 \pm 0.013$  & $0.049 \pm 0.009$   & $0.061 \pm 0.008$ \\
&StyleGAN 2          &               &               & $0.701 \pm 0.009$    & $0.659 \pm 0.002$  & $0.404 \pm 0.006$   & $0.588 \pm 0.006$ \\
&StyleGAN 3          &               &               & $0.701 \pm 0.006$    & $0.675 \pm 0.003$  & $0.380 \pm 0.003$   & $0.585 \pm 0.004$ \\
&Diffusion           &               &               & $0.734 \pm 0.010$    & $0.727 \pm 0.002$  & $0.533 \pm 0.004$   & $0.664 \pm 0.005$ \\
    \end{tabular}
    \vspace{0.2cm}
    \caption{Results on the test dataset (56 subjects) when training the generative models with BraTS 2020 (313 training subjects). Mean and standard deviation of Dice score across the dataset are calculated for the labels; GD-enhancing tumor (ET), peritumoral edema (ED), and necrotic and non-enhancing tumor core (NCR/NET). All results are presented as mean $\pm$ standard deviation of 10 trainings. The column Orig marks if the original dataset has been added to the training set and the column Aug marks if augmentation was used during training of the segmentation network. Training with and without augmentation was performed with the same number of total images.} The top row shows baseline results using only real images. 
    \label{tab:res_ens_brats2020}
\end{table*}

\begin{table*}  \scriptsize
    \centering
    \begin{tabular}{c|c|c|c|ccc|c} 
    
&        \textbf{Model} & \textbf{Orig} & \textbf{Aug} & \textbf{ET} & \textbf{ED} & \textbf{NCR/NET} & \textbf{Mean} \\ \hline

\multirow{22}{*}{\rotatebox{90}{\textbf{U-Net}}}
&None & \checkmark & \checkmark & $0.847 \pm 0.010$ & $0.829 \pm 0.003$ & $0.709 \pm 0.011$ & $0.795 \pm 0.006$ \\
&Progressive GAN & \checkmark & \checkmark & $0.821 \pm 0.056$ & $0.813 \pm 0.038$ & $0.688 \pm 0.035$ & $0.774 \pm 0.042$ \\
&\textbf{StyleGAN 1} & \checkmark & \checkmark & $\bf{0.853 \pm 0.007}$ & $\bf{0.828 \pm 0.004}$ & $\bf{0.704 \pm 0.016}$ & $\bf{0.795 \pm 0.006}$ \\
&StyleGAN 2 & \checkmark & \checkmark & $0.780 \pm 0.201$ & $0.757 \pm 0.228$ & $0.647 \pm 0.200$ & $0.728 \pm 0.209$ \\
&StyleGAN 3 & \checkmark & \checkmark & $0.843 \pm 0.014$ & $0.825 \pm 0.016$ & $0.717 \pm 0.020$ & $0.795 \pm 0.014$ \\
&Diffusion & \checkmark & \checkmark & $0.835 \pm 0.011$ & $0.821 \pm 0.006$ & $0.704 \pm 0.013$ & $0.787 \pm 0.007$ \\
&None & \checkmark & & $0.845 \pm 0.009$ & $0.828 \pm 0.003$ & $0.708 \pm 0.012$ & $0.794 \pm 0.006$ \\
&Progressive GAN & \checkmark & & $0.830 \pm 0.041$ & $0.821 \pm 0.028$ & $0.692 \pm 0.028$ & $0.781 \pm 0.031$ \\
&StyleGAN 1 & \checkmark & & $0.818 \pm 0.121$ & $0.792 \pm 0.161$ & $0.669 \pm 0.151$ & $0.760 \pm 0.144$ \\
&StyleGAN 2 & \checkmark & & $0.810 \pm 0.145$ & $0.794 \pm 0.165$ & $0.675 \pm 0.145$ & $0.759 \pm 0.151$ \\
&StyleGAN 3 & \checkmark & & $0.839 \pm 0.015$ & $0.826 \pm 0.016$ & $0.708 \pm 0.021$ & $0.791 \pm 0.015$ \\
&Diffusion & \checkmark & & $0.840 \pm 0.011$ & $0.824 \pm 0.005$ & $0.699 \pm 0.014$ & $0.788 \pm 0.007$ \\
&Progressive GAN & &  \checkmark & $0.638 \pm 0.032$ & $0.582 \pm 0.035$ & $0.525 \pm 0.026$ & $0.582 \pm 0.029$ \\
&StyleGAN 1 & &  \checkmark & $0.710 \pm 0.011$ & $0.682 \pm 0.007$ & $0.377 \pm 0.015$ & $0.590 \pm 0.007$ \\
&StyleGAN 2 & &  \checkmark & $0.734 \pm 0.022$ & $0.738 \pm 0.016$ & $0.606 \pm 0.024$ & $0.692 \pm 0.015$ \\
&StyleGAN 3 & &  \checkmark & $0.746 \pm 0.013$ & $0.738 \pm 0.004$ & $0.473 \pm 0.016$ & $0.652 \pm 0.005$ \\    
&Diffusion & &  \checkmark & $0.757 \pm 0.009$ & $0.743 \pm 0.010$ & $0.632 \pm 0.017$ & $0.711 \pm 0.009$ \\
&Progressive GAN & & & $0.516 \pm 0.142$ & $0.400 \pm 0.185$ & $0.344 \pm 0.184$ & $0.420 \pm 0.165$ \\
&StyleGAN 1 & & & $0.635 \pm 0.085$ & $0.585 \pm 0.103$ & $0.282 \pm 0.098$ & $0.501 \pm 0.091$ \\
&StyleGAN 2 & & & $0.584 \pm 0.156$ & $0.571 \pm 0.170$ & $0.367 \pm 0.240$ & $0.507 \pm 0.186$ \\
&StyleGAN 3 & & & $0.725 \pm 0.024$ & $0.719 \pm 0.020$ & $0.445 \pm 0.031$ & $0.629 \pm 0.024$ \\
&\textbf{Diffusion} & & & $\bf{0.772 \pm 0.018}$ & $\bf{0.764 \pm 0.023}$ & $\bf{0.646 \pm 0.020}$ & $\bf{0.727 \pm 0.019}$ \\ \hline
\multirow{22}{*}{\rotatebox{90}{\textbf{Swin Transformer}}}
&None                & \checkmark    & \checkmark    & $0.779 \pm 0.009$    & $0.816 \pm 0.005$  & $0.674 \pm 0.011$   & $0.756 \pm 0.008$ \\
&Progressive GAN     & \checkmark    & \checkmark    & $0.785 \pm 0.012$    & $0.812 \pm 0.002$  & $0.664 \pm 0.010$   & $0.754 \pm 0.009$ \\
&StyleGAN 1          & \checkmark    & \checkmark    & $0.777 \pm 0.008$    & $0.817 \pm 0.003$  & $0.664 \pm 0.008$   & $0.752 \pm 0.006$ \\
&StyleGAN 2          & \checkmark    & \checkmark    & $0.790 \pm 0.006$    & $0.813 \pm 0.002$  & $0.666 \pm 0.012$   & $0.756 \pm 0.006$ \\
&\textbf{StyleGAN 3  }        & \checkmark    & \checkmark    & $\textbf{0.795} \pm \textbf{0.005}$    & $\textbf{0.813} \pm \textbf{0.003}$  & $\textbf{0.671} \pm \textbf{0.010}$   & $\textbf{0.760} \pm \textbf{0.006}$ \\
&\textbf{Diffusion}           & \checkmark    & \checkmark    & $\textbf{0.791} \pm \textbf{0.007}$    & $\textbf{0.816} \pm \textbf{0.003}$  & $\textbf{0.672} \pm \textbf{0.009}$   & $\textbf{0.760} \pm \textbf{0.006}$ \\
&None                & \checkmark    &               & $0.772 \pm 0.002$    & $0.811 \pm 0.002$  & $0.661 \pm 0.010$   & $0.748 \pm 0.005$ \\
&Progressive GAN     & \checkmark    &               & $0.781 \pm 0.008$    & $0.810 \pm 0.002$  & $0.659 \pm 0.014$   & $0.750 \pm 0.008$ \\
&StyleGAN 1          & \checkmark    &               & $0.778 \pm 0.008$    & $0.810 \pm 0.004$  & $0.650 \pm 0.013$   & $0.746 \pm 0.008$ \\
&StyleGAN 2          & \checkmark    &               & $0.786 \pm 0.010$    & $0.809 \pm 0.003$  & $0.656 \pm 0.011$   & $0.750 \pm 0.008$ \\
&StyleGAN 3          & \checkmark    &               & $0.789 \pm 0.009$    & $0.810 \pm 0.003$  & $0.668 \pm 0.015$   & $0.756 \pm 0.009$ \\
&Diffusion           & \checkmark    &               & $0.788 \pm 0.010$    & $0.814 \pm 0.002$  & $0.663 \pm 0.013$   & $0.755 \pm 0.008$ \\
&Progressive GAN     &               & \checkmark    & $0.691 \pm 0.009$    & $0.672 \pm 0.002$  & $0.470 \pm 0.003$   & $0.611 \pm 0.004$ \\
&StyleGAN 1          &               & \checkmark    & $0.720 \pm 0.002$    & $0.696 \pm 0.002$  & $0.375 \pm 0.004$   & $0.597 \pm 0.003$ \\
&StyleGAN 2          &               & \checkmark    & $0.716 \pm 0.007$    & $0.739 \pm 0.001$  & $0.501 \pm 0.003$   & $0.652 \pm 0.004$ \\
&StyleGAN 3          &               & \checkmark    & $0.731 \pm 0.002$    & $0.744 \pm 0.002$  & $0.429 \pm 0.002$   & $0.635 \pm 0.002$ \\
&\textbf{Diffusion}           &               & \checkmark    & $\textbf{0.767} \pm \textbf{0.003}$    & $\textbf{0.794} \pm \textbf{0.002}$  & $\textbf{0.628} \pm \textbf{0.010}$   & $\textbf{0.729} \pm \textbf{0.005}$ \\
&Progressive GAN     &               &               & $0.659 \pm 0.004$    & $0.636 \pm 0.005$  & $0.461 \pm 0.004$   & $0.585 \pm 0.004$ \\
&StyleGAN 1          &               &               & $0.703 \pm 0.007$    & $0.683 \pm 0.003$  & $0.356 \pm 0.003$   & $0.581 \pm 0.004$ \\
&StyleGAN 2          &               &               & $0.710 \pm 0.002$    & $0.736 \pm 0.002$  & $0.500 \pm 0.002$   & $0.649 \pm 0.002$ \\
&StyleGAN 3          &               &               & $0.715 \pm 0.009$    & $0.732 \pm 0.002$  & $0.421 \pm 0.001$   & $0.623 \pm 0.004$ \\
&Diffusion           &               &               & $0.768 \pm 0.003$    & $0.789 \pm 0.003$  & $0.627 \pm 0.010$   & $0.728 \pm 0.005$ \\
                
    \end{tabular}
    \vspace{0.2cm}
    \caption{Results on the test dataset (56 subjects) when training the generative models with BraTS 2021 (1195 training subjects). Mean and standard deviation of Dice score across the dataset are calculated for the labels; GD-enhancing tumor (ET), peritumoral edema (ED), and necrotic and non-enhancing tumor core (NCR/NET). All results are presented as mean $\pm$ standard deviation of 10 trainings. The column Orig marks if the original dataset has been added to the training set and the column Aug marks if augmentation was used during training of the segmentation network. Training with and without augmentation was performed with the same number of total images.} The top row shows baseline results using only real images.
    \label{tab:res_ens_brats2021}
\end{table*}

\begin{table*}  \scriptsize
    \centering
    \begin{tabular}{c|c|c|c|c|c} 
    
&        \textbf{Model} & \textbf{Orig} & \textbf{Aug} & \textbf{Relative Dice BraTS 2020} &  \textbf{Relative Dice BraTS 2021}\\ \hline
\multirow{22}{*}{\rotatebox{90}{\textbf{U-Net}}}
&Progressive GAN & \checkmark & \checkmark & 100.14\% & 97.36\% \\
&StyleGAN 1 & \checkmark &\checkmark       & 96.98\% & \textbf{100.00\%} \\
&StyleGAN 2 & \checkmark & \checkmark      & \textbf{101.10\%} & 91.57\% \\
&StyleGAN 3 & \checkmark &\checkmark       & 95.74\% & \textbf{100.00\%} \\
&Diffusion & \checkmark &\checkmark        & 100.55\% & 98.99\%  \\
&Progressive GAN & \checkmark &            & 99.59\% & 98.24\% \\
&StyleGAN 1 & \checkmark &                 & 98.49\% & 95.60\%  \\
&StyleGAN 2 & \checkmark &                 & 100.41\% & 95.47\%\\
&StyleGAN 3 & \checkmark &                 & 97.80\% &  99.5\%\\
&Diffusion  & \checkmark &                 & 100.27\% & 99.12\% \\

&Progressive GAN & & \checkmark            & 79.70\% & 73.21\% \\
&StyleGAN 1 &  &\checkmark                 & 12.21\% & 74.21\% \\
&StyleGAN 2 &  & \checkmark                & 93.00\% & 87.04\% \\
&StyleGAN 3 &  &\checkmark                 & 87.80\% & 82.01\% \\
&Diffusion &  &\checkmark                  & \textbf{100.14\%} & \textbf{89.43\%}  \\
&Progressive GAN & &                       & 62.27\% & 52.83\% \\
&StyleGAN 1 & &                            & 9.47\% & 63.02\%  \\
&StyleGAN 2 & &                            & 65.71\% & 63.77\%\\
&StyleGAN 3 & &                            & 81.62\% & 79.12\%\\
&Diffusion  & &                            & 99.18\% & 91.45\% \\ \hline

\multirow{22}{*}{\rotatebox{90}{\textbf{Swin Transformer}}}
&Progressive GAN & \checkmark & \checkmark & \textbf{95.42\%} & 99.73\% \\
&StyleGAN 1 & \checkmark &\checkmark       & 92.87\% & 99.47\% \\
&StyleGAN 2 & \checkmark & \checkmark      & 94.48\% & 100.00\% \\
&StyleGAN 3 & \checkmark &\checkmark       & 93.27\% & \textbf{100.52\%}\\
&Diffusion & \checkmark &\checkmark        & 93.41\% & \textbf{100.52\%} \\
&Progressive GAN & \checkmark &            & 89.5\%  & 99.20\% \\
&StyleGAN 1 & \checkmark &                 & 90.04\% & 98.68\% \\
&StyleGAN 2 & \checkmark &                 & 91.11\%  &  99.20\% \\
&StyleGAN 3 & \checkmark &                 & 91.65\% & 100.0\% \\
&Diffusion  & \checkmark &                 & 91.11\% & 99.87\% \\

&Progressive GAN & & \checkmark            & 81.69\% & 80.82\% \\
&StyleGAN 1 &  &\checkmark                 & 8.74\%  & 78.97\% \\
&StyleGAN 2 &  & \checkmark                & 84.66\% & 86.24\% \\
&StyleGAN 3 &  &\checkmark                 & 81.16\% & 83.99\% \\
&Diffusion &  &\checkmark                  & 92.06\% & \textbf{96.42\%} \\
&Progressive GAN & &                       & 78.73\%  & 77.38\% \\
&StyleGAN 1 & &                            & 8.21\%  & 76.85\% \\
&StyleGAN 2 & &                            & 79.13\% & 85.84\% \\
&StyleGAN 3 & &                            & 78.73\%  & 82.41\% \\
&Diffusion  & &                            & \textbf{89.37\%} & 96.30\% \\
    \end{tabular}
    \vspace{0.2cm}
    \caption{Results on the test datasets (56 subjects) when training the generative models with BraTS 2020 and 2021. Relative Dice is defined as the mean Dice score when using only synthetic images, or real and synthetic, divided by the mean Dice score when using only real images with augmentation. Aug marks if augmentation was used during training of the segmentation network. Except for StyleGAN 1, the relative Dice scores are lower for BraTS 2021 when only using synthetic images. Training with and without augmentation was performed with the same number of total images. }
    \label{tab:relative_dice}
\end{table*}

\clearpage

\subsection{Hausdorff distance}

Tables~\ref{tab:res_ens_brats2020_haus} and~\ref{tab:res_ens_brats2021_haus} in the Appendix show obtained Hausdorff distances when training the segmentation networks with synthetic images from BraTS 2020 and 2021, respectively. Overall the rankings of the generative models are very similar to the ranking from the Dice scores, the main difference being that the progressive GAN is ranked higher for U-Net. The mean Hausdorff distance is in general much lower (i.e. better) for the Swin transformer compared to the U-Net.

\subsection{Qualitative evaluation by neuroradiologist}

In addition to the quantitative metrics a qualitative evaluation by an experienced neuroradiologist was performed, see the Methods section for details. Table~\ref{tab:qualitative} shows the results of the evaluation, i.e. how the images were classified (real or synthetic).

\begin{table*}  \scriptsize
    \centering
\begin{tabular}{c|c|c}
\textbf{Image type}          & \textbf{Classified as} real & \textbf{Classified as} synthetic \\
\hline
Real image     & \textbf{16}   & 84         \\
Progressive GAN & 23   & \textbf{77}         \\
StyleGAN 1       & 14   & \textbf{86}        \\
StyleGAN 2       & 13   & \textbf{87}        \\
StyleGAN 3       & 18   & \textbf{82}        \\
Diffusion Model & 16   & \textbf{84}       
\end{tabular}%
    \vspace{0.2cm}
    \caption{Results from the qualitative evaluation by an experienced neuroradiologist, who classified 600 four-panel images as real or synthetic (100 real images and 100 images per generative model). The two right columns show how many images that were classified as real or synthetic, for real images and for synthetic images from the five generative models. The number of correct classifications for each row has been marked with bold.}
    \label{tab:qualitative}
\end{table*}

\clearpage
\section{Discussion}

Our evaluation shows that training segmentation networks with synthetic images works well, with Dice scores that reach 91\% - 100\% compared to when training with real images (for BraTS 2021 and 2020, respectively). Shin et al.~\cite{shin2018medical} obtained a relative Dice score of 77.6\% for brain tumor segmentation, using the deep convolutional GAN architecture which is older than progressive GAN. Fernandez et al.~\cite{fernandez2022can} obtained a similar relative Dice of 93.8\%, but only performed a binary tumor segmentation which is an easier task. No comparison with other generative models was conducted. Thambawita et al.~\cite{thambawita2022singan} obtained a relative Dice score of 97.2\%, but for segmentation of endoscopy images making it difficult to compare the results. Furthermore, the authors did not use more recent generative models like StyleGAN or diffusion models.

The Dice scores for the diffusion model are for BraTS 2020 basically the same as when training with real images, which made us suspicious. An investigation revealed that the diffusion model had memorized many of the training images~\cite{akbar2023beware}. Memorization has previously been shown when using diffusion models for non-medical images~\cite{somepalli2023diffusion,carlini2023extracting}, but to the best of our knowledge not for medical images. Diffusion models are more likely to memorize the training images compared to GANs~\cite{carlini2023extracting,akbar2023beware}, due to a completely different architecture.

Even better results can be obtained using an ensemble of 5 - 10 generative models~\cite{eilertsen,dikici2021constrained,larsson2022does}, as each model by random chance will learn a different subset of the high dimensional distribution, at the cost of a training time which is 5 - 10 times longer. Larsson et al.~\cite{larsson2022does} demonstrated that using an ensemble of 10 progressive GANs improved the mean Dice score for brain tumor segmentation by 9.5\%, compared to a single GAN. The benefit of using an ensemble is expected to be larger for BraTS 2021, compared to BraTS 2020 used in~\cite{larsson2022does}, to capture all modes of the distribution (due to a larger number of imaging sites in BraTS 2021). 

Training the generative models with 1195 subjects (BraTS 2021), instead of 313 (BraTS 2020), leads to worse performance for the U-Net (lower relative Dice scores when using only synthetic images, mean 83.68\% versus 77.36\%, excluding StyleGAN 1) which may seem surprising. However, BraTS 2021 contains data from a larger number of sites (23 versus 19), which will result in more modes in the high dimensional distribution, which is harder to learn. Furthermore, using a larger dataset like BraTS 2021 makes it harder for the generative models to memorize the training images~\cite{somepalli2023diffusion,akbar2023beware}. It would be very interesting to compare our results to SinGAN-Seg~\cite{thambawita2022singan}, where a single image is used to train the generative model, but we suspect that such a model is prone to memorization.


Our results show that augmentation makes a rather big difference for the U-Net when training with only synthetic images, while the improvement is smaller for the Swin transformer. A possible explanation for this is that the augmentation helps the segmentation networks to overcome systematic differences between real and synthetic images. To apply augmentation when training the generative models needs to be explored in future work, as it on the one hand can increase the number of training images, but on the other hand it may introduce more modes in the high dimensional distribution (which will be harder to learn).

Several other researchers have demonstrated that combining real and synthetic images (i.e. using generative models for advanced augmentation) can improve segmentation accuracy~\cite{shin2018medical,bowles2018gan,pollastri2020augmenting}, or classification accuracy~\cite{frid2018gan,azizi2023synthetic}, compared to training with only real images, but our results show that adding synthetic images only provides minor improvements or even results in worse performance. There are at least two possible explanations for this. First, we use a rather strong baseline segmentation model with several types of traditional augmentation during training. Second, we repeat the training of each segmentation network 10 times to avoid differences due to random chance. It is possible that using a subset of the real data (e.g. 20\%), instead of all the real data, would result in larger improvements when adding synthetic images. However, the generative models should then be trained with the same subset, which will reduce the quality and diversity of the synthetic images.

The results are likely to strongly depend on the hyperparameters of each generative model, but to explore many parameter combinations is difficult due to the long training times of both generative models and segmentation networks. The total training time for this work was over 2000 days on an Nvidia A100 graphics card. This work demonstrates that new efficient metrics for evaluating synthetic medical images are required, as FID and IS are based on ImageNet (which does not contain medical images), do not consider memorization~\cite{somepalli2023diffusion,carlini2023extracting,akbar2023beware}, and do in general not correlate with how a network trained on the synthetic images will perform (see Table~\ref{tab:ranking})~\cite{barratt2018note,eilertsen}. In future work we will calculate FID and IS using CNNs pre-trained on RadImageNet~\cite{mei2022radimagenet}, which is a large collection of medical images, to see if Rad-FID and Rad-IS better correlate with our other metrics.

Regarding the qualitative evaluation by a neuroradiologist, the results show that the generative models produce synthetic images that are on the same level as real images (a similar number of images were classified as synthetic). It should however be noted that the setup of this experiment was not similar to a regular clinical assessment of brain tumor MRI, which is reflected in the fact that a large portion of the real images were falsely classified as synthetic images. Normally, in a clinical workflow a neuroradiologist would assess the whole brain, instead of a single slice, with even more MRI sequences than used in this study. Furthermore, a neuroradiologist does not normally look at skullstripped brain images. A challenge with this evaluation was that the BraTS images originate from many different scanners, with varying image quality, which probably also affected the visual assessment. Two additional limitations are that only one neuroradiologist performed the visual assessment, and that the sample of 600 images is not balanced in terms of real and synthetic images (which may introduce a bias).

The implication of our results are that sharing synthetic medical images is a viable option to sharing real images. A researcher can use synthetic images for pre-training, and then fine-tune the model on a small number of locally available images. Sharing synthetic medical images can be substantially easier~\cite{rankin2020reliability,el2021evaluating,rajotte2022synthetic,thambawita2022singan}, as GDPR should not apply for data which do not belong to a specific person (but further legal research is needed). Regarding  consent, Larson et al.~\cite{larson2020ethics} argue that clinical data should be treated as a form of public good, to be used for the benefit of future patients, and further argue that consent is not required before collected data are used for secondary purposes when obtaining such consent is prohibitively costly or burdensome (e.g. contacting 1,000 - 10,000 persons). On the other hand, the argument of clinical data being treated as a form of public good, and not requiring further consent for use in research or development, may be a slippery slope and many examples exist where a retrospective look identifies the continued use of such data as an unauthorized abuse (e.g. Henrietta Lacks).

Before sharing synthetic images it is important to investigate how similar each synthetic image is to all training images, as especially diffusion models have been shown to memorize the training images~\cite{somepalli2023diffusion,carlini2023extracting,akbar2023beware,dar2023investigating}. This is extra important for small datasets, as memorization is then more likely~\cite{somepalli2023diffusion,akbar2023beware}. Common evaluation metrics like FID and IS do not capture memorization, and it is therefore necessary to for example calculate the correlation, or some other metric like mutual information, between each synthetic image and all training images~\cite{dikici2021constrained,akbar2023beware,dar2023investigating}. Pre-trained generative models can play an important role for sharing synthetic images from small datasets, as it should be less likely for a pre-trained model to memorize a small number of new images during fine tuning (compared to training the model from scratch). To determine an acceptable range of overlap with real clinical data is a very difficult task, especially since different legal experts interpret GDPR differently (in Sweden it is in general interpreted stricter compared to other countries) and since this acceptable range is likely to be different for different types of medical images. It therefore remains an open question how high the highest similarity can be before a synthetic image is seen as a copy of a training image.

\section{Methods}

\subsection{Data}

The MR images used for this project were downloaded from the Multimodal Brain tumour Segmentation Challenge (BraTS) 2020 and 2021~\cite{bakas3,bakas4,bakas1,bakas2,menze,baid2021rsna}. The training set contains MR volumes of shape 240 × 240 × 155 from 369 subjects for BraTS 2020 and from 1251 subjects for BraTS 2021. For each subject four types of MR images are available: T1-weighted (T1w), post gadolinium contrast T1-weighted (T1wGd), T2-weighted (T2w), and T2-weighted fluid attenuated inversion recovery (FLAIR). The annotations cover three parts of the brain tumor: peritumoural edema (ED), necrotic and non-enhancing tumour core (NCR/NET), and GD-enhancing tumour (ET). We used 313/1195 subjects for training and 56 subjects for testing, after first performing a random shuffling of the subjects. The data in the test sets were not  used for training the generative models.

All 3D volumes were split into 2D slices, as a 2D GAN and a 2D diffusion model were used (3D GANs and 3D diffusion models are not yet very common). Only slices with at least 15\% pixels with an intensity of more than 50 were included in the training. This resulted in a total of 23,478 5-channel images for BraTS 2020, and 91,271 5-channel images for BraTS 2021. Each slice was zero padded from 240 x 240 to 256 x 256 pixels, as the used GANs only work for resolutions that are a power of 2, and the intensity was rescaled to 0 - 255. The intensities for the tumor annotations were changed from [1,2,4] to [51,102,204], such that the intensity range is more similar for the 5 channels. 

\subsection{Image generation}

In this work we compare four different GANs (progressive growing GAN~\cite{karras2018progressive}, StyleGAN 1~\cite{karras2019style}, StyleGAN 2~\cite{karras2020analyzing}, StyleGAN 3~\cite{karras2021alias}) and a diffusion model~\cite{ho2020denoising,nichol2021improved}, for the task of generating brain tumor images. GANs are trained through adversarial learning (using an adversarial loss function), where a generator and a discriminator compete against each other, to produce more realistic images and to be better at discriminating images as real or synthetic. At inference time, only the generator is used. A diffusion model, on the other hand, starts with real data samples, progressively adds noise over many steps according to a predetermined schedule until the data becomes pure noise. The diffusion model is then trained to reverse this process, using more traditional loss functions, reconstructing less noisy data from more noisy data at each step. During inference, a diffusion model starts with an image of pure noise and sequentially applies the learned denoiser to reduce the noise, following the reverse of the training noise schedule. This process is iterated until the noise is completely removed, resulting in the generation of a new image which resembles the training data distribution. In general diffusion models are easier to train compared to GANs, due to more traditional loss functions, but are much slower at generating images.

The openly available code of each generative model was modified to generate 5-channel images instead of 3 channels, no other modifications to the default architectures were done. Each generative model will thereby learn to jointly generate the four MR images (T1w, T1wGD, T2w, FLAIR) and the corresponding tumor annotation at the same time. There is no guarantee that the synthetic annotations will
be restricted to the same values as the real annotations
([51,102,204]). The synthetic annotations were therefore
thresholded to the closest original annotation value.

The used hyperparameters of each model, and the approximate training times, are provided in Table~\ref{tab:models_parameters}. We used a set of common hyperparameters across all models, along with some model-specific ones. For instance, in the case of StyleGANs, we experimented with different gamma values, the best of which are detailed in the accompanying table. For the diffusion models, they were trained with varying diffusion steps, but the optimal results were obtained with 4000 steps.   

For each generative model a total of 100,000 synthetic 5-channel images were generated. For the GANs this took about 10 minutes, while it took 1.5 days (using 8 GPUs) for the diffusion model The synthetic images and the trained generative models are shared on AIDA data hub~\cite{hedlund2020key,dataset}.

\begin{table*}  \scriptsize
    \centering
    \begin{tabular}{c|c|c|c|c|c} 
    
        \textbf{Model} & \textbf{Batch size} & \textbf{Iterations} & \textbf{gamma} & \textbf{Hardware} &  \textbf{Training time} \\ \hline
Progressive GAN & 128 - 8 & 12000 & - & 1 x Nvidia V100 & 5 days  \\
StyleGAN 1 & 16 & 25000 & 10 & 2 x Nvidia V100 & 6 days  \\
StyleGAN 2 & 32 & 25000 & 8.2 & 4 x Nvidia A100 & 3 days  \\
StyleGAN 3 & 32 & 25000 & 8.2 & 4 x Nvidia A100 & 3 days  \\
Diffusion model & 12 & 7800 & - & 1 x Nvidia V100 & 22 days  \\

    \end{tabular}
    \vspace{0.2cm}
    \caption{The used hyperparameters for the different generative models, as well as the used hardware and the approximate training time. One iteration corresponds to 1000 images processed. Processing 1000 images for the diffusion model is much slower, since it performs 4000 diffusion steps for each batch.}
    \label{tab:models_parameters}
\end{table*}

\subsection{Quantitative evaluation and tumor segmentation}

The quality and diversity of synthetic images are often evaluated using metrics such as Frechét inception distance (FID) and inception score (IS)~\cite{borji2019pros}. Since these metrics are based on CNNs trained on ImageNet, which does not contain medical images, they will be biased for medical images. Furthermore, these metrics will not tell us how well a network trained with synthetic images will perform on real images. The synthetic images were therefore used to train segmentation networks (based on U-Net and Swin transformers), and the evaluation was performed using real images in the test set. To investigate how FID and IS correlate with the performance of training with only synthetic images, FID and IS were also calculated. 


Training deep networks is a stochastic process, meaning that training the same model several times will give different results. Each segmentation network was therefore trained 10 times, to make sure that performance differences between the different generative models are not due to random chance. The segmentation was performed for each slice independently, and the Dice scores and Hausdorff distances were then calculated in 3D after putting the slices for each subject back into a volume.

\subsubsection{U-Net}

The model structure and training setup used was inspired by the 2D segmentation code from nnUNet~\cite{isensee2019automated}. The model was a U-Net~\cite{ronneberger} with an extra depth layer and instance normalization instead of batch normalization. In addition, the ReLU activations was swapped for leaky ReLUs with a negative slope of $10^{-2}$. All models were trained with a loss consisting of a cross-entropy term and a soft Dice term weighted equally. In addition, deep supervision was used, meaning that the loss was applied on the five highest depth level with weighting $0.5^d$ where $d$ is the depth.

The loss was minimized using stochastic gradient descent with Nesterov momentum of $0.99$ and weight decay of $3 \cdot 10^{-5}$. The initial learning rate was $5 \cdot 10^{-2}$ and was decreased using polynomial learning rate decay with an exponent of $0.9$. The learning rate and optimizer was different for the generative models, see their respective paper. All models were trained for $3 \cdot 10^7$ samples or 3 days, whichever occured first. 20\% of the available images were used for validation and the model with the best mean Dice over the validation set was used for evaluation. If both real and synthetic data was used during training, the real dataset and the synthetic dataset were sampled equally often.

During training a geometric, and intensity-augmentation was applied, as our previous work on augmentation for brain tumor segmentation~\cite{cirillo2021best} demonstrated that augmentation can provide significant improvements even if the dataset is large. The image and target is first randomly rotated and scaled. Both rotation and scaling is applied with a probability of 0.75, the image and target is rotated with an angle uniformly sampled from $[-30^{\circ}, 30^{\circ}]$ and the width and height is scaled (independently) with a scale factor uniformly sampled from $[0.9, 1.1]$. Then the four input channels are augmented by; adding Gaussian noise (applied with probability 0.5, zero mean and standard deviation uniformly sampled from $[0.0, 0.05]$), blurring the image (applied with probability 0.2, Gaussian blurring with standard deviation sampled from $[0.5, 1.0]$), faking lower resolution (zooming with a factor between 0.75 and 1.0 and then upsampling) and changing the gamma factor (scaling it with a factor between 0.8 and 1.2). Lastly the input image channels are normalized using Z-score normalization.

\subsubsection{Swin transformers} The Swin transformer segmentation network was implemented and trained using the MMSegmentation library \cite{MMSegmentation}. The architecture employed is the Swin-Base variant, as implemented in MMSegmentation, with a window size of 7 and a patch size of 4 x 4. The original Swin transformer was designed for 3-channel RGB images, hence, to accommodate MRI scans with four modalities per slice, the number of input channels in the model was modified to 4. Consequently, the input dimension of the Swin transformer is set to 256 x 256 x 4, where the '4' denotes the number of channels, and '256' represents both the height and width of each modality slice. The sole alteration to the original Swin-Base architecture is the adaptation of both the number of channels and the number of classes to 4.

The network was trained using DiceCELoss function. Furthermore, the Dice Loss component does not take into account the background class, label 0. The loss per batch was derived by calculating the loss for each training image, and taking the mean loss value.

The AdamW optimizer with $\beta$1 = 0.9, $\beta$2 = 0.999 and weight decay $\lambda$ = 0.01 was used. Additionally, a learning rate scheduler with warmup and linear decay was employed. The warmup ratio was set to $1e^{-6}$. For the first 1500 iterations, the learning rate is increased and after 1500 warmup iterations, the learning rate reached 0.00006. For the rest of the training, the learning rate was decreased linearly until it reached 0.0 at the end of the training. The models were trained for 25 epochs, and after every epoch, a validation loss was calculated. The batch size was set to 8, and the last batch is dropped if it is not the same size as all of the other batches to ensure that all models are provided with batches of consistent size. 

Data augmentation was performed using five techniques inspired by Eklund et al.~\cite{cirillo2021best}. Images and their corresponding segmentations undergo rotation, with angles from $0^\circ$ to $30^\circ$ chosen randomly, and scaling, with axis factors varying within $\pm20\%$. Images were also subjected to a 50\% chance of either horizontal or vertical flipping. Brightness was adjusted through a power-law $\gamma$ intensity transformation, with parameters randomly picked between 0.8 and 1.2. Lastly, elastic deformation, following the methodology from the original U-Net paper~\cite{ronneberger2015u}, was applied using a deformation grid with normal distribution displacements ($\sigma$ = 2 voxels) and smoothed with a third-order spline filter.

\subsection{Qualitative evaluation by neuroradiologist}

\begin{figure}[htb]
  \centering
  \includegraphics[width=0.49\textwidth]{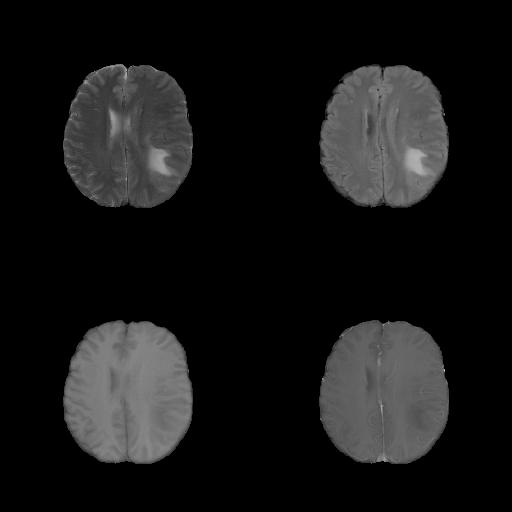} 
  \includegraphics[width=0.49\textwidth]{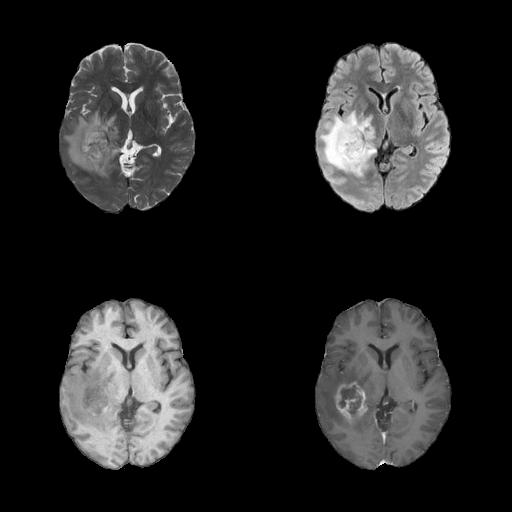} 
  \caption{Left: a real 4-channel image shown during the qualitative evaluation, where the task was to classify each example as real or synthetic. Right: a synthetic 4-channel image shown during the qualitative evaluation.}
  \label{fig:Turing_sample}
\end{figure}

To evaluate how the synthetic images are perceived by a clinician, a total of 600 4-sequence-panel images (T1w, T1wGd, T2w, T2w FLAIR) were presented to an experienced (13 years) neuroradiologist (co-author IB). The task was to determine if each presented 4-panel image was real or synthetic. The 600 images consisted of 100 real images, and 500 synthetic images, 100 each from the five generative models (progressive GAN, StyleGAN 1-3, diffusion model). The total number of real images, and the number of synthetic images per generative model, was known to the neuroradiologist before starting the evaluation. The real and synthetic images were presented in a random order. The evaluation took approximately 12 hours. Figure~\ref{fig:Turing_sample} shows an example of a real and a synthetic image presented during the evaluation.

\section{Data availability}

The BraTS 2020 and 2021 datasets are openly available through the following websites. \\

\noindent https://www.med.upenn.edu/cbica/brats2020/data.html \\ 

\noindent http://braintumorsegmentation.org \\

\noindent The generated synthetic images (100,000 five channel images per generative model), and the trained generative models, are shared at the AIDA data hub~\cite{hedlund2020key,dataset}; https://datahub.aida.scilifelab.se/10.23698/aida/synthetic/brgandi 

\section{Code availability}

The code for the different GANs and the diffusion model is openly shared by the creators of the generative models, see below. We therefore share our modifications to make the code work for 5-channel images, instead of 3-channel images, the used segmentation code, and some additional help scripts. \\

\noindent https://github.com/muhamadusman/Assist/ \\

\noindent Generative models \\

\noindent Progressive growing GAN, https://github.com/tkarras/progressive\_growing\_of\_gans \\

\noindent StyleGAN 1, https://github.com/NVlabs/stylegan \\

\noindent StyleGAN 2, https://github.com/NVlabs/stylegan2 \\

\noindent StyleGAN 3, https://github.com/NVlabs/stylegan3 \\

\noindent Diffusion model, https://github.com/openai/guided-diffusion \\

\noindent Segmentation models \\

\noindent U-Net, https://github.com/MIC-DKFZ/nnUNet\\

\noindent Swin Transformer, https://github.com/open-mmlab/mmsegmentation

\section*{Ethics}

This research study was conducted retrospectively using anonymized human subject data made available by BraTS. The ethical review board of Link\"{o}ping decided that no further ethical approval was required.

\section*{Contributions}

MUA trained all the generative models, the Swin transformer segmentation networks, generated synthetic images and calculated FID and IS metrics. ML trained all the U-Net segmentation networks and calculated the segmentation metrics. IB performed the qualitative evaluation. AE drafted the manuscript and contributed with conceptualization, supervision and funding. All authors contributed to the interpretation of the results, have revised and edited the manuscript and approved the submitted version.

\section*{Competing interests}

AE has previously received graphics hardware from Nvidia. The other authors declare no competing interests.

\section{Acknowledgements}

Training several of the generative models and all the segmentation networks was performed using the supercomputing resource Berzelius (752 Nvidia A100 GPUs) provided by the National Supercomputer Centre at Linköping University, Sweden. It was donated by the Knut and Alice Wallenberg foundation. 

This research was supported by the ITEA/VINNOVA project ASSIST (Automation, Surgery Support and Intuitive 3D visualization to optimize workflow in IGT SysTems, 2021-01954), LiU Cancer and the Åke Wiberg foundation. Ida Blystad was supported by a research grant from the Wallenberg Center for Molecular Medicine as an associated clinical fellow.

\bibliographystyle{splncs04}
\bibliography{refrences}

\appendix
\section{Appendix}

\begin{figure*}
    \setlength{\tabcolsep}{0pt} 
    \begin{tabular}{cccccccc}
       
        & Annotation & None & StyleGAN 1 & StyleGAN 2 & StyleGAN 3 & Progressive GAN & Diffusion \\
        \rotatebox{90}{ Orig + Aug} &
        \includegraphics[width=0.16\textwidth]{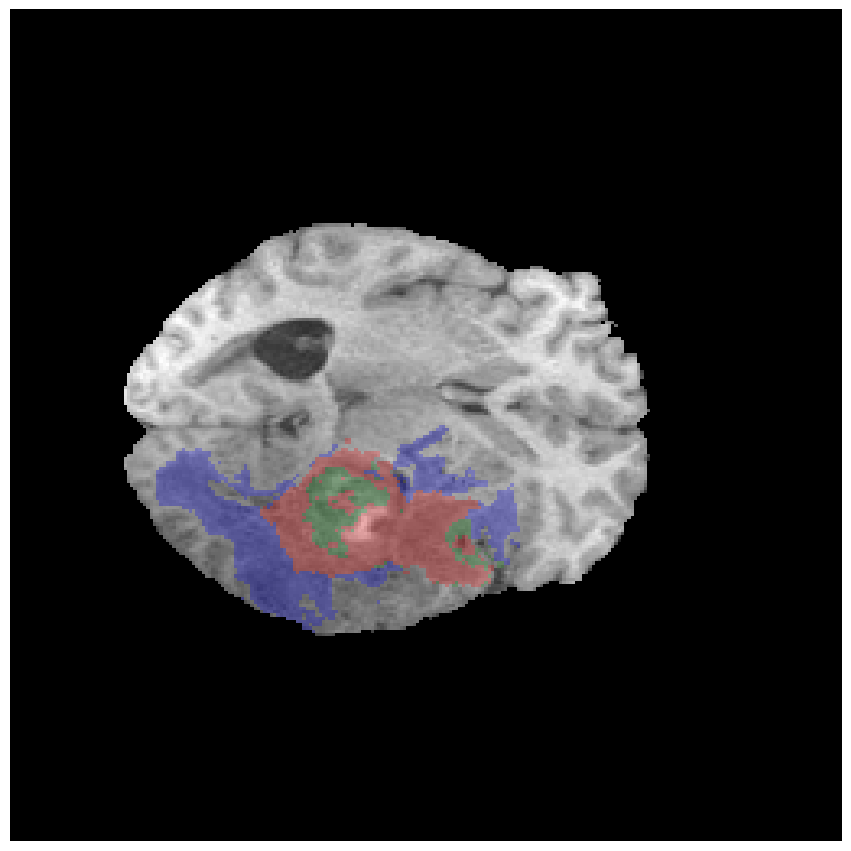} &
        \includegraphics[width=0.16\textwidth]{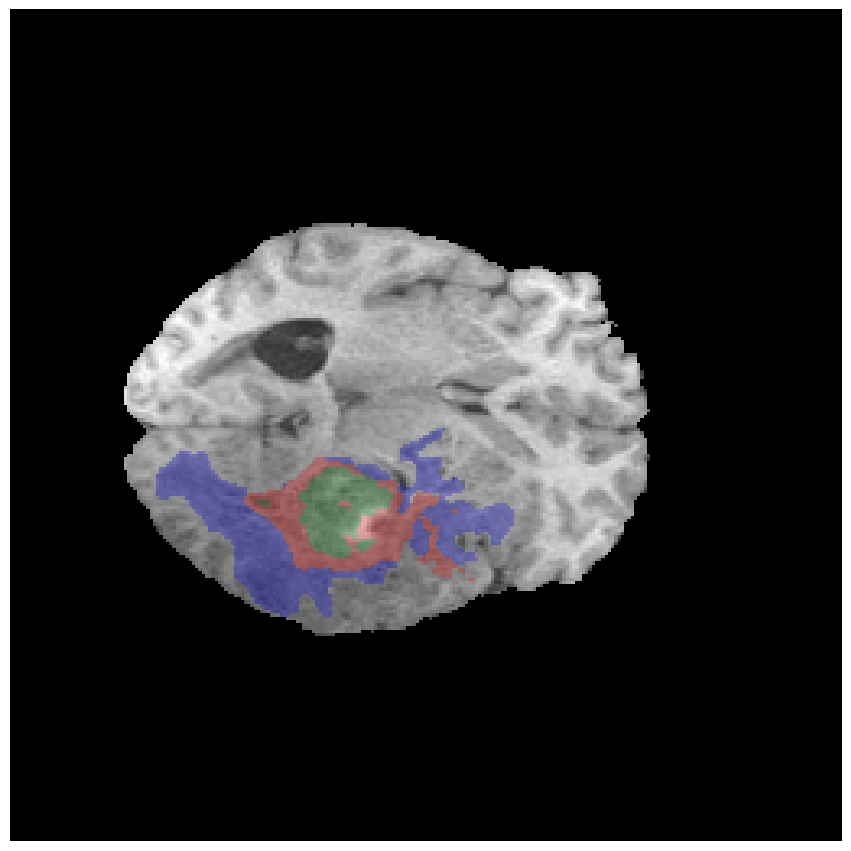} &
        \includegraphics[width=0.16\textwidth]{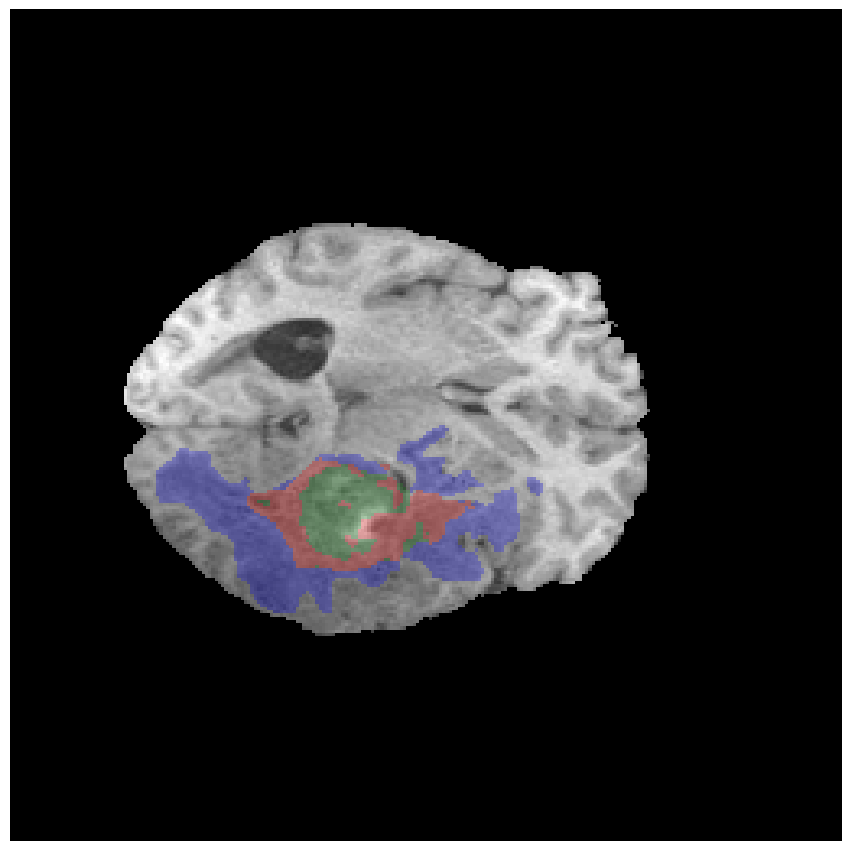} &
        \includegraphics[width=0.16\textwidth]{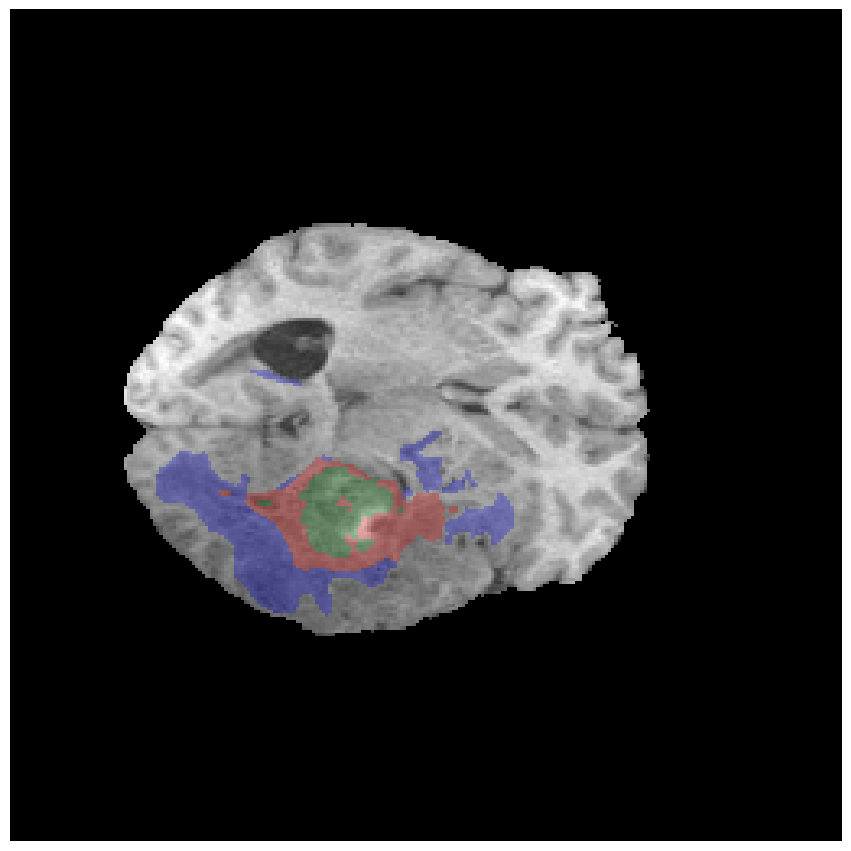} &
        \includegraphics[width=0.16\textwidth]{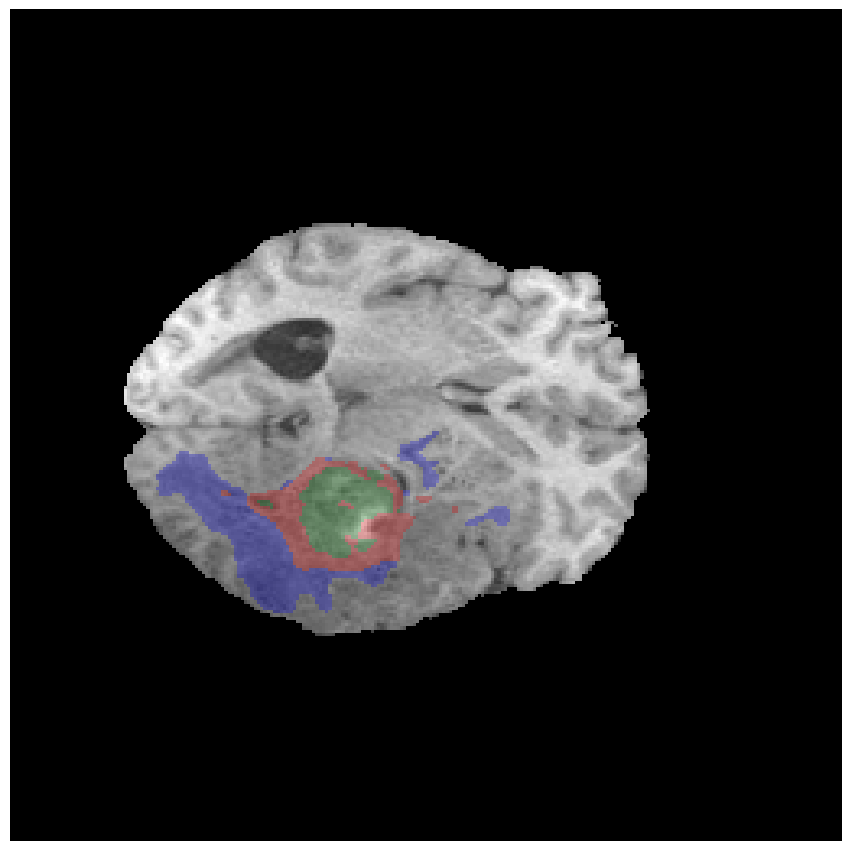} &
        \includegraphics[width=0.16\textwidth]{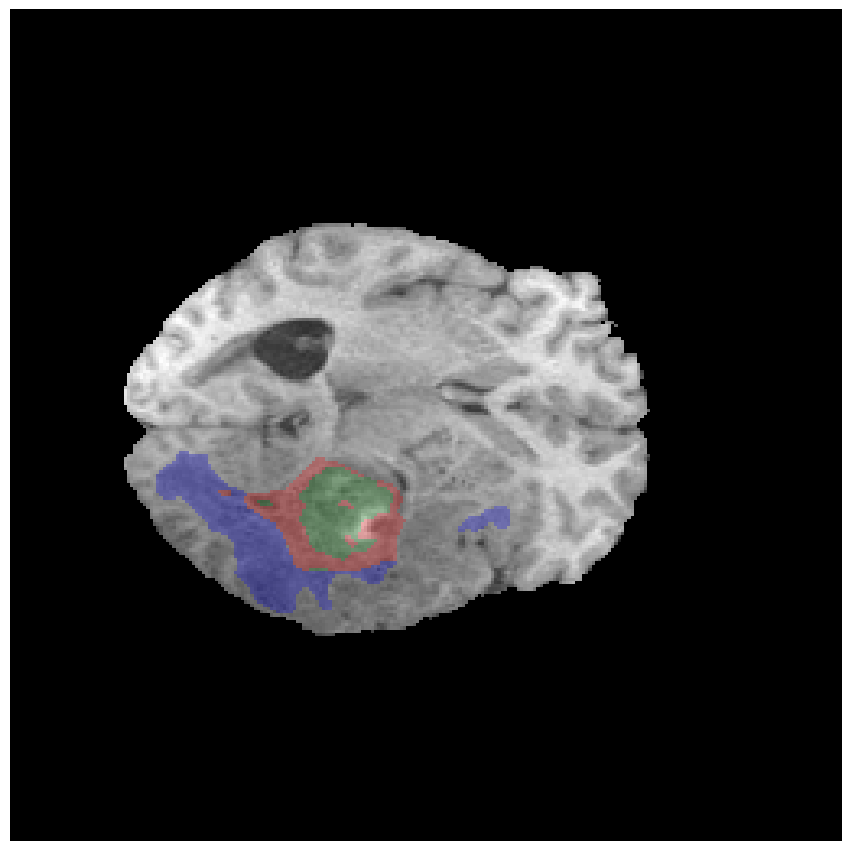} &
        \includegraphics[width=0.16\textwidth]{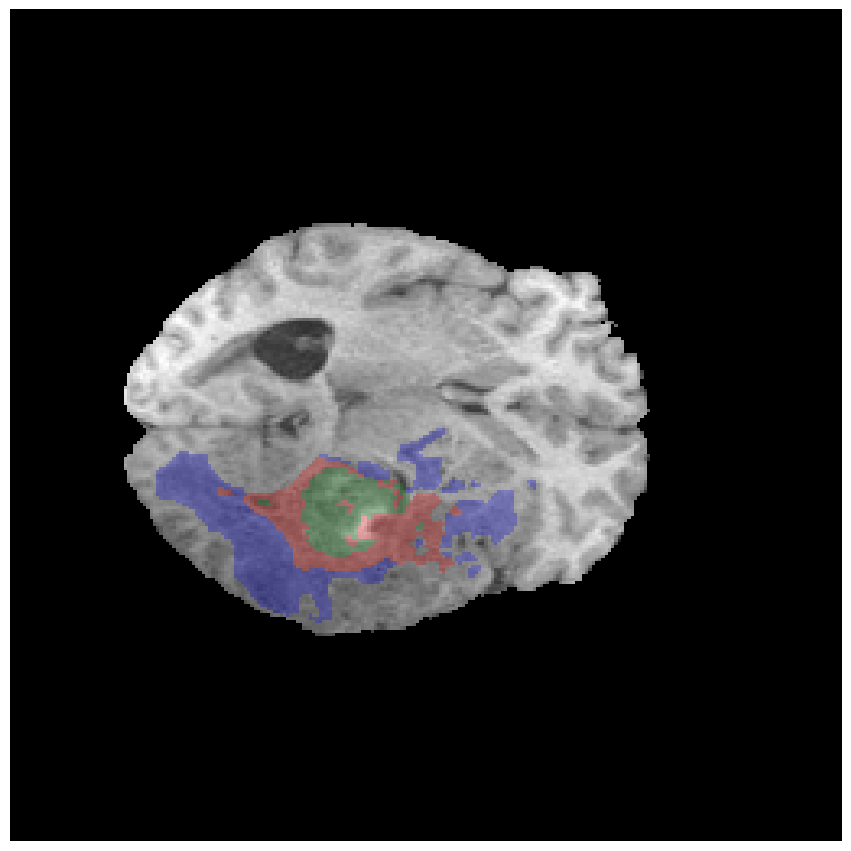} \\
        
        \rotatebox{90}{\hspace{.5cm} Orig} &
        \includegraphics[width=0.16\textwidth]{Figures/results/313/Subject_009_slice_76_T1/gt.png} &
        \includegraphics[width=0.16\textwidth]{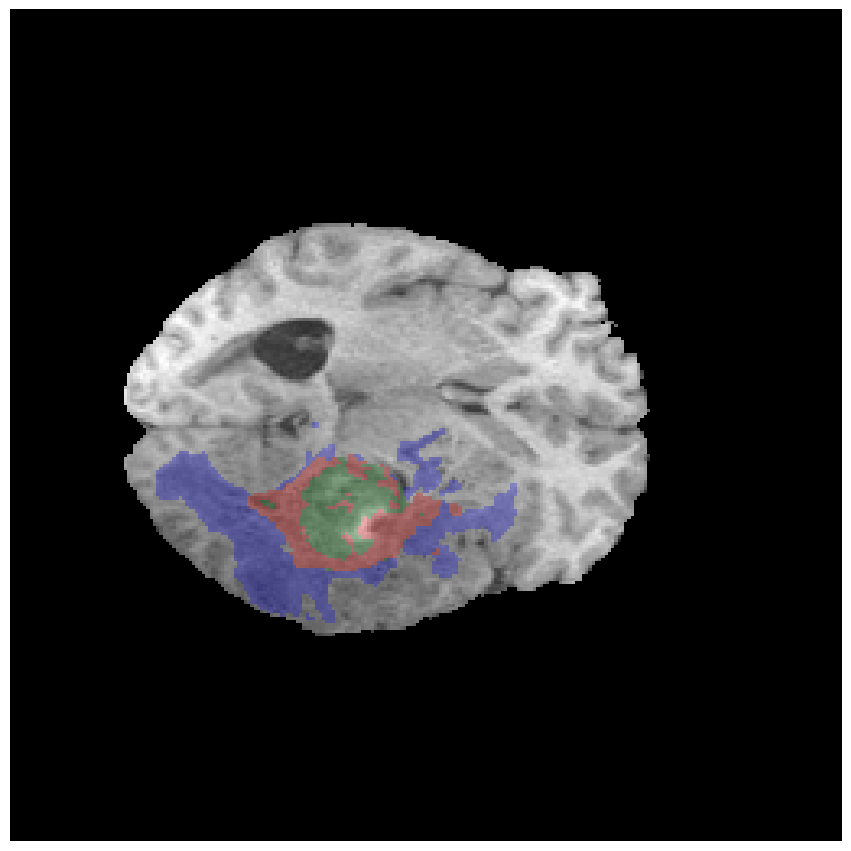} &
        \includegraphics[width=0.16\textwidth]{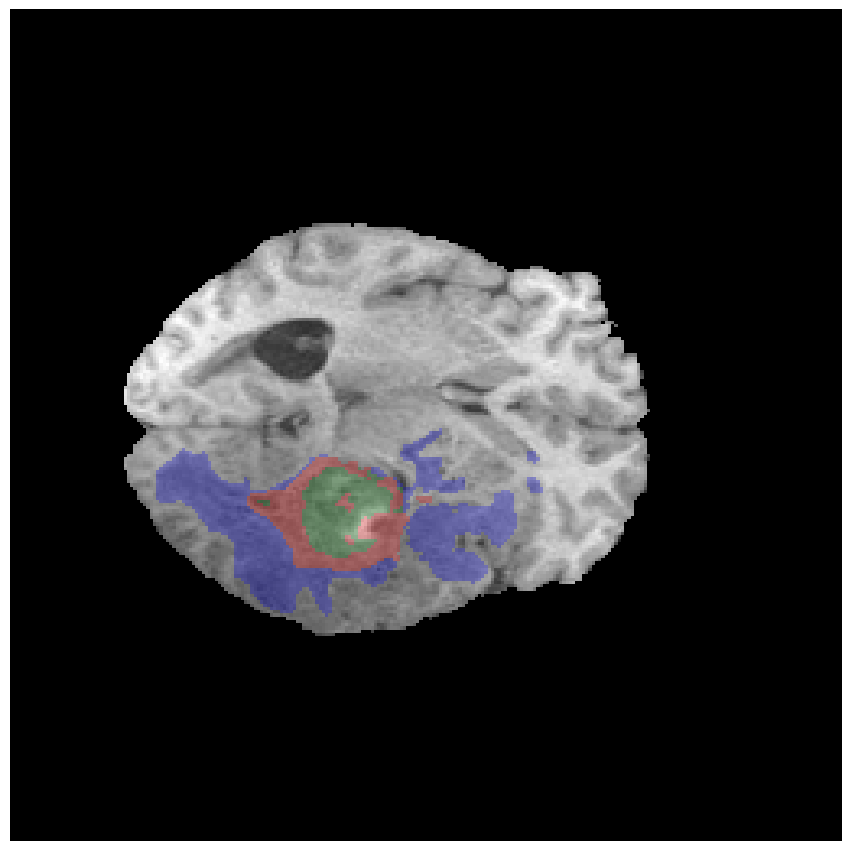} &
        \includegraphics[width=0.16\textwidth]{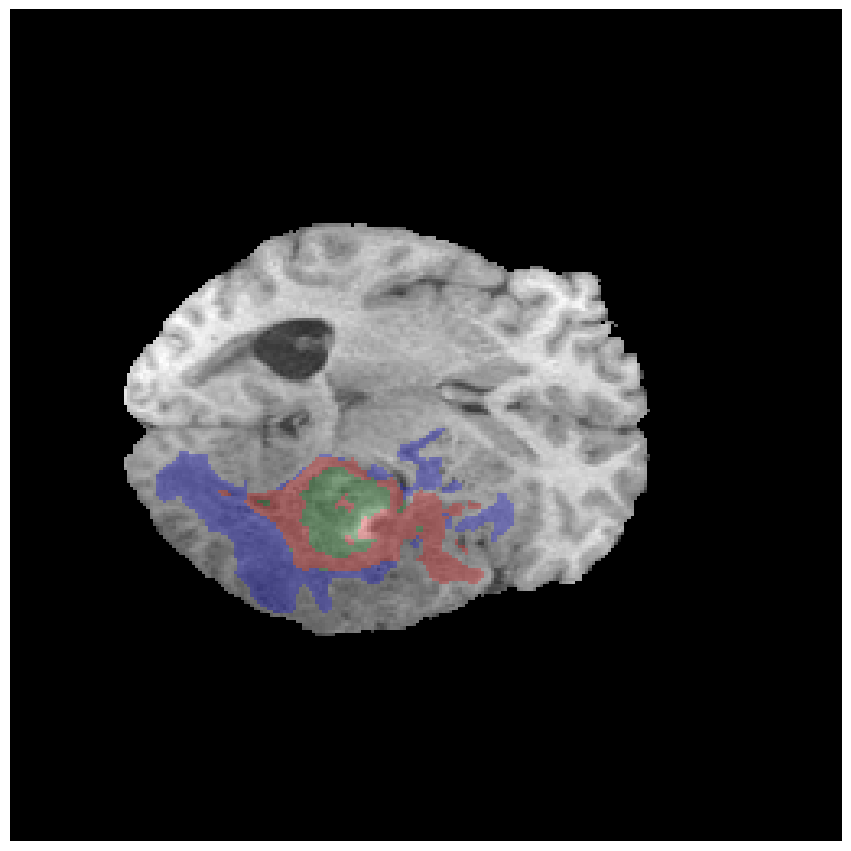} &
        \includegraphics[width=0.16\textwidth]{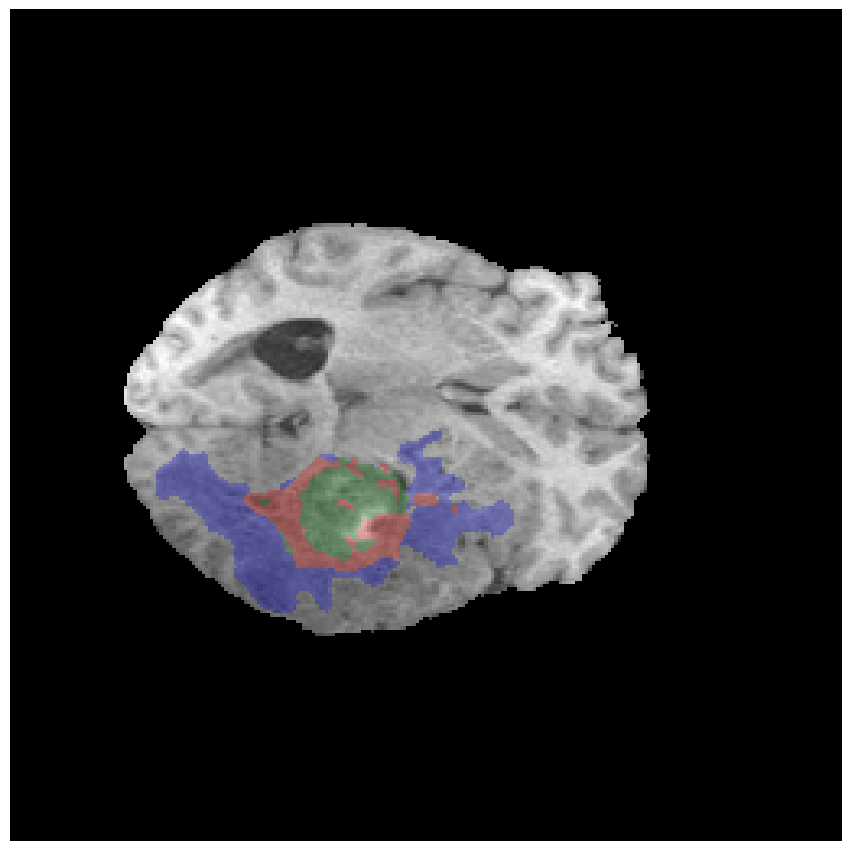} &
        \includegraphics[width=0.16\textwidth]{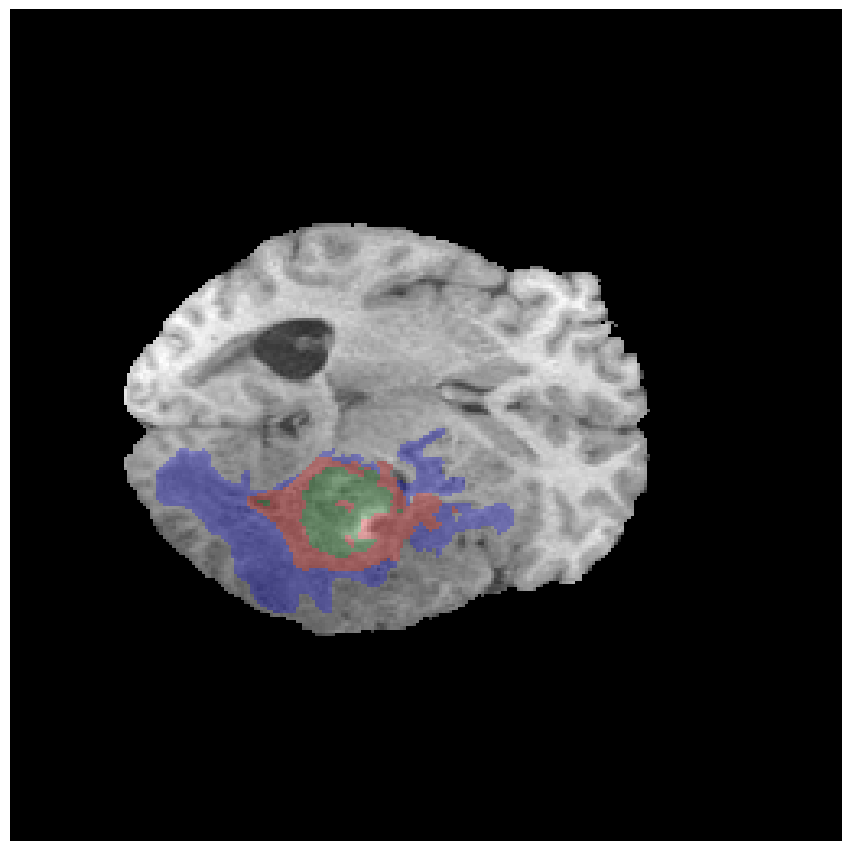} &
        \includegraphics[width=0.16\textwidth]{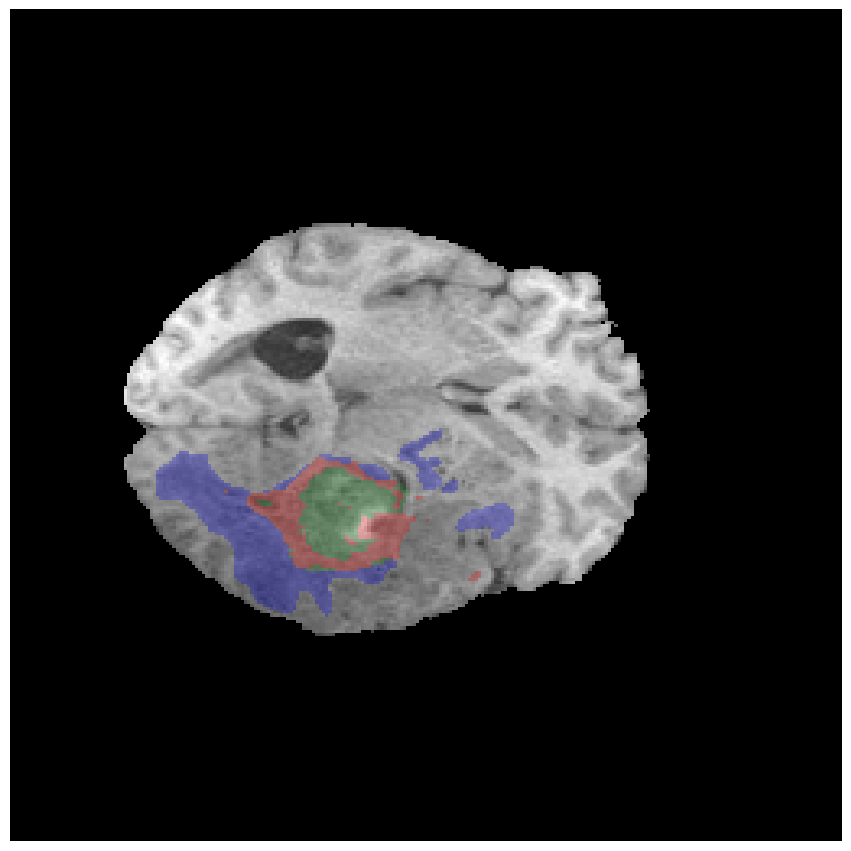} \\

        \rotatebox{90}{\hspace{.5cm} Aug} &
        \includegraphics[width=0.16\textwidth]{Figures/results/313/Subject_009_slice_76_T1/gt.png} &
         &
        \includegraphics[width=0.16\textwidth]{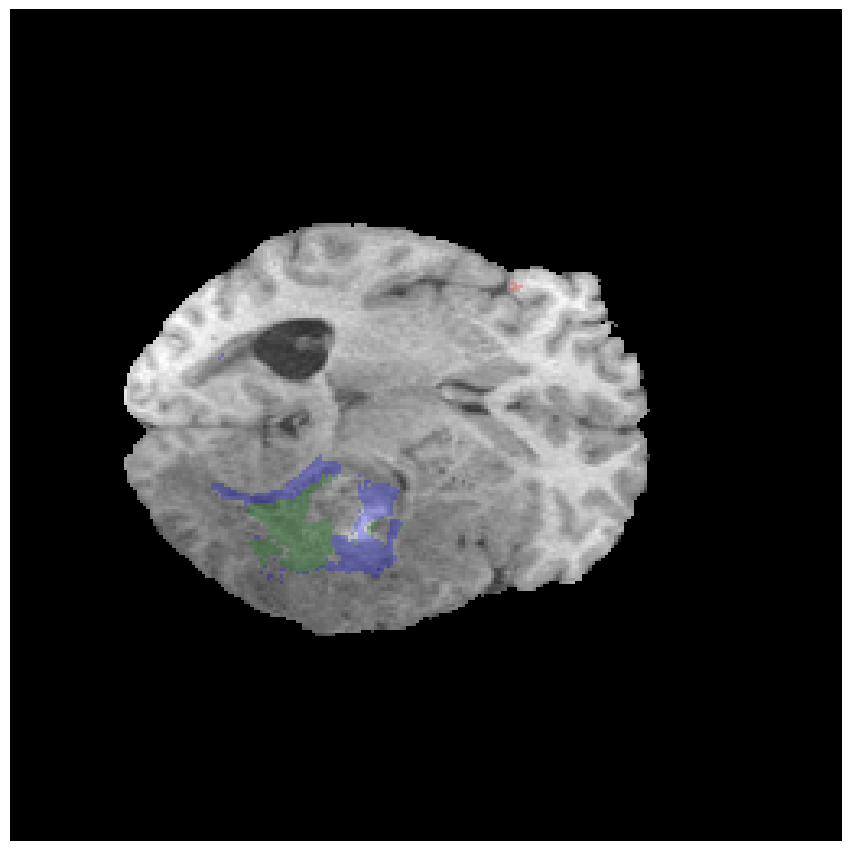} &
        \includegraphics[width=0.16\textwidth]{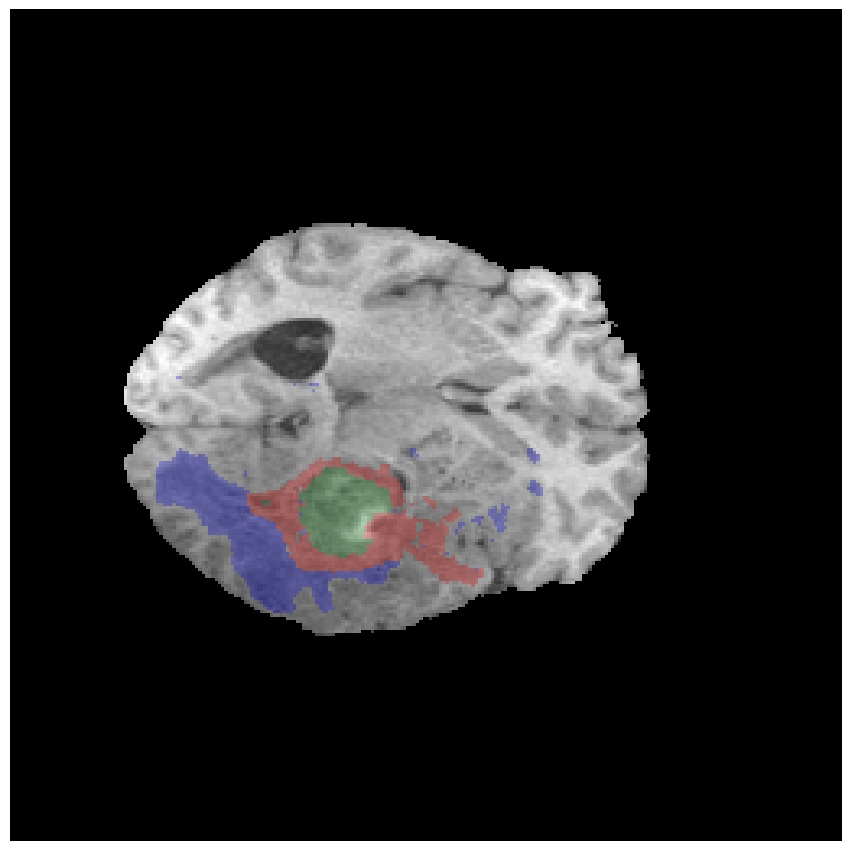} &
        \includegraphics[width=0.16\textwidth]{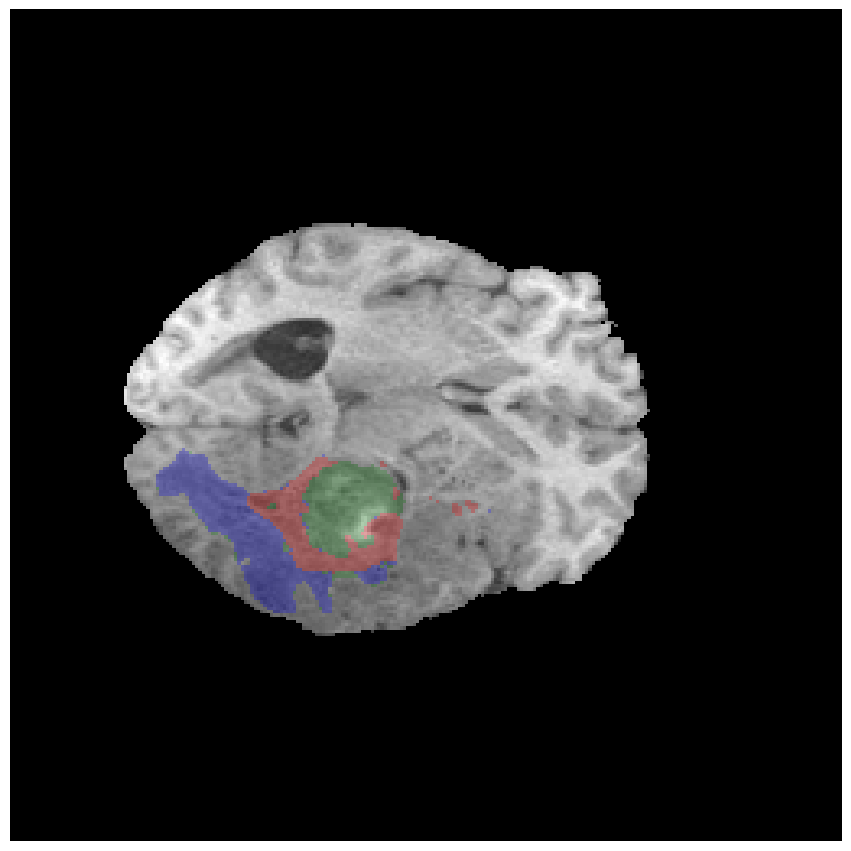} &
        \includegraphics[width=0.16\textwidth]{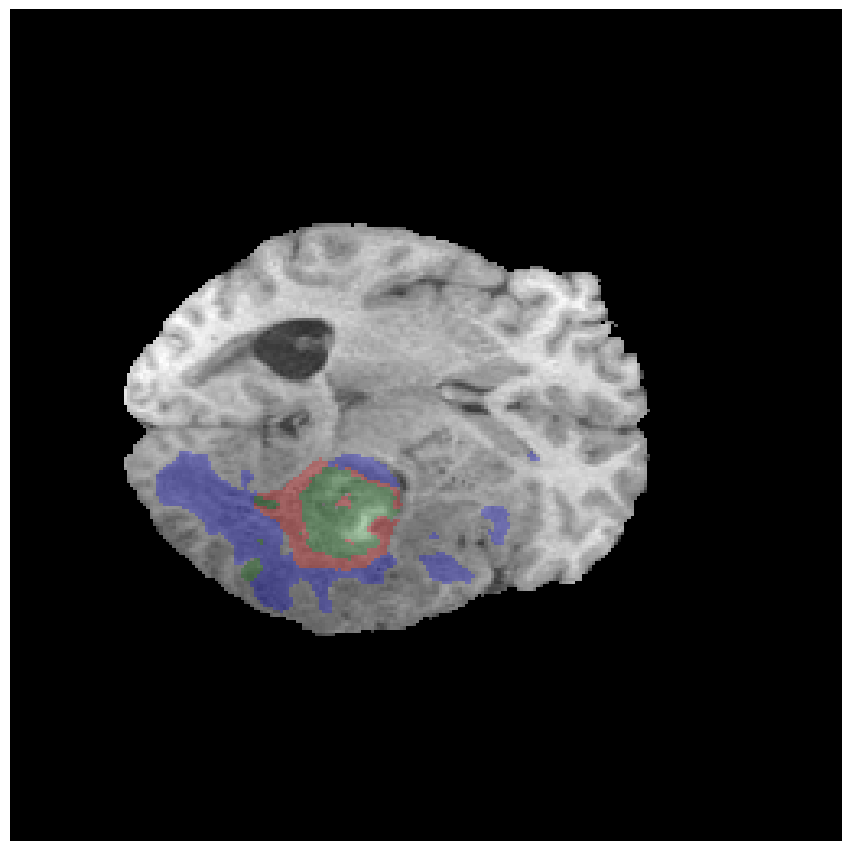} &
        \includegraphics[width=0.16\textwidth]{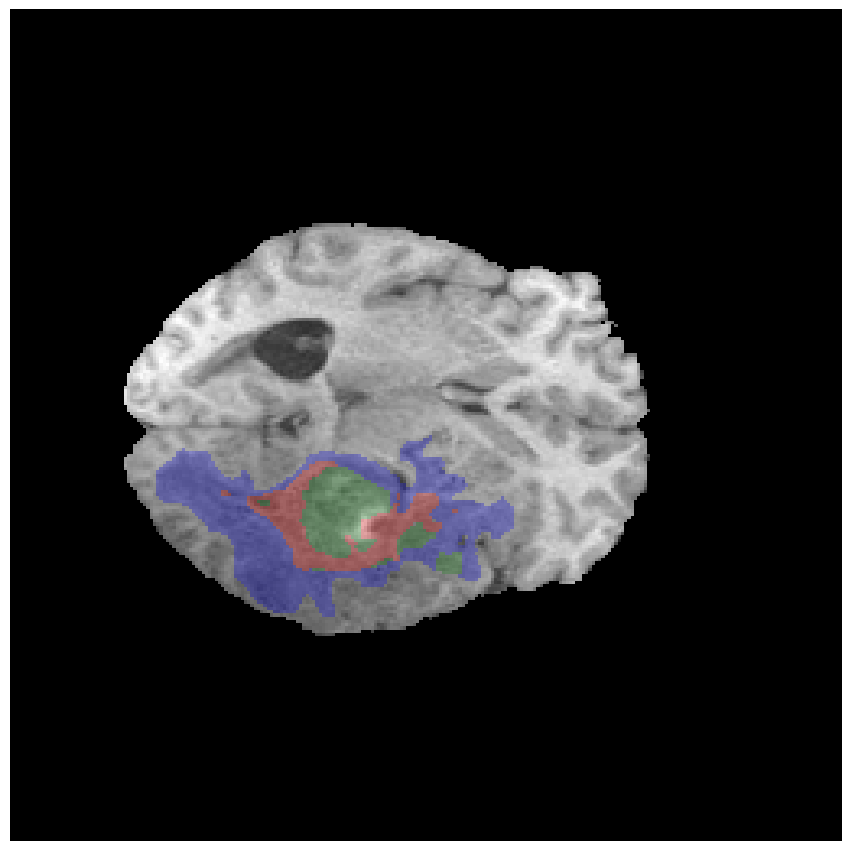} \\

        &
        \includegraphics[width=0.16\textwidth]{Figures/results/313/Subject_009_slice_76_T1/gt.png} &
         &
        \includegraphics[width=0.16\textwidth]{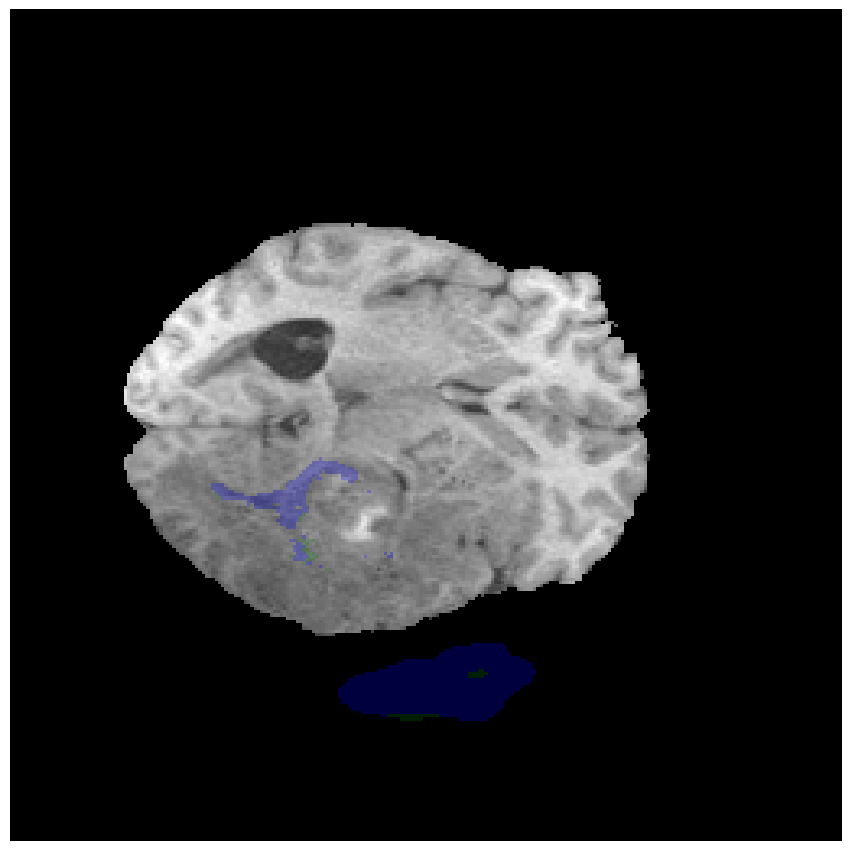} &
        \includegraphics[width=0.16\textwidth]{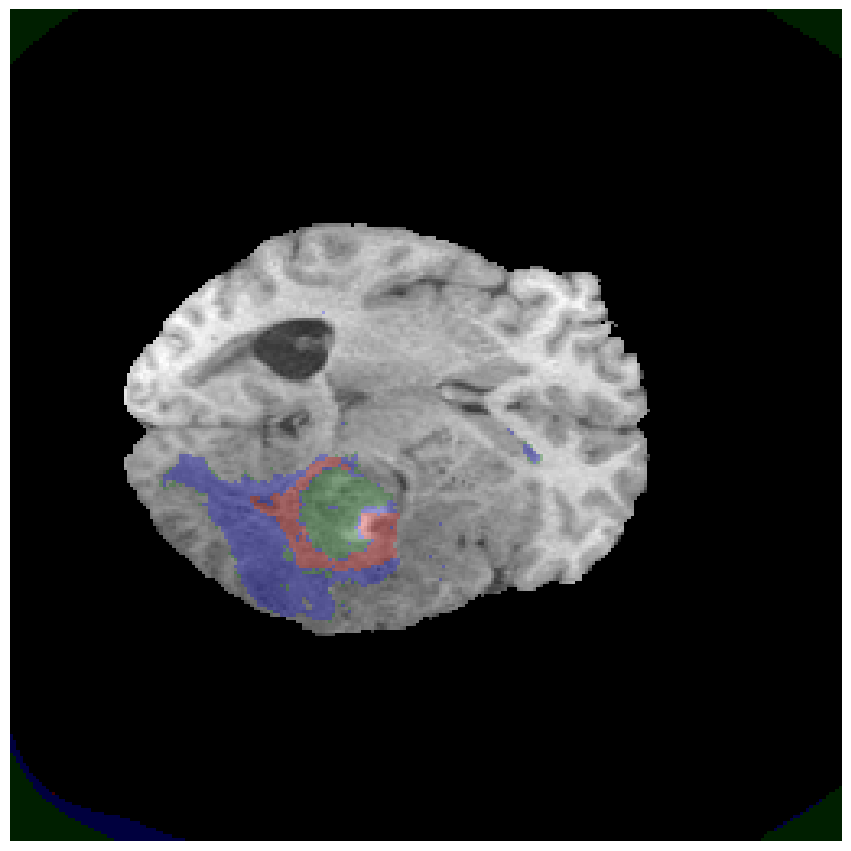} &
        \includegraphics[width=0.16\textwidth]{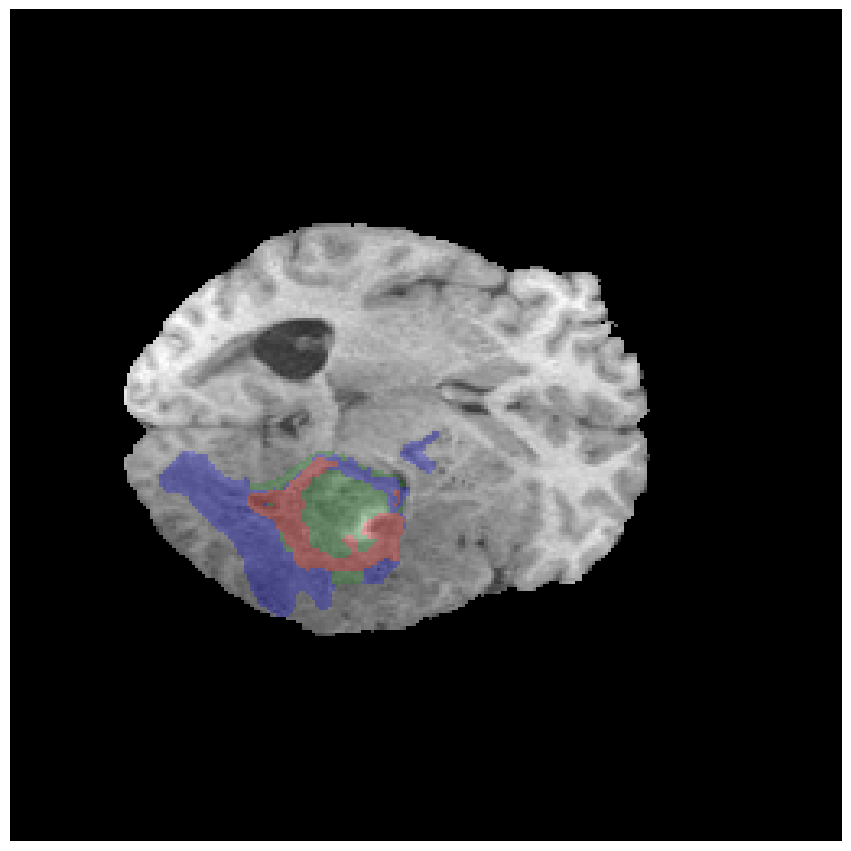} &
        \includegraphics[width=0.16\textwidth]{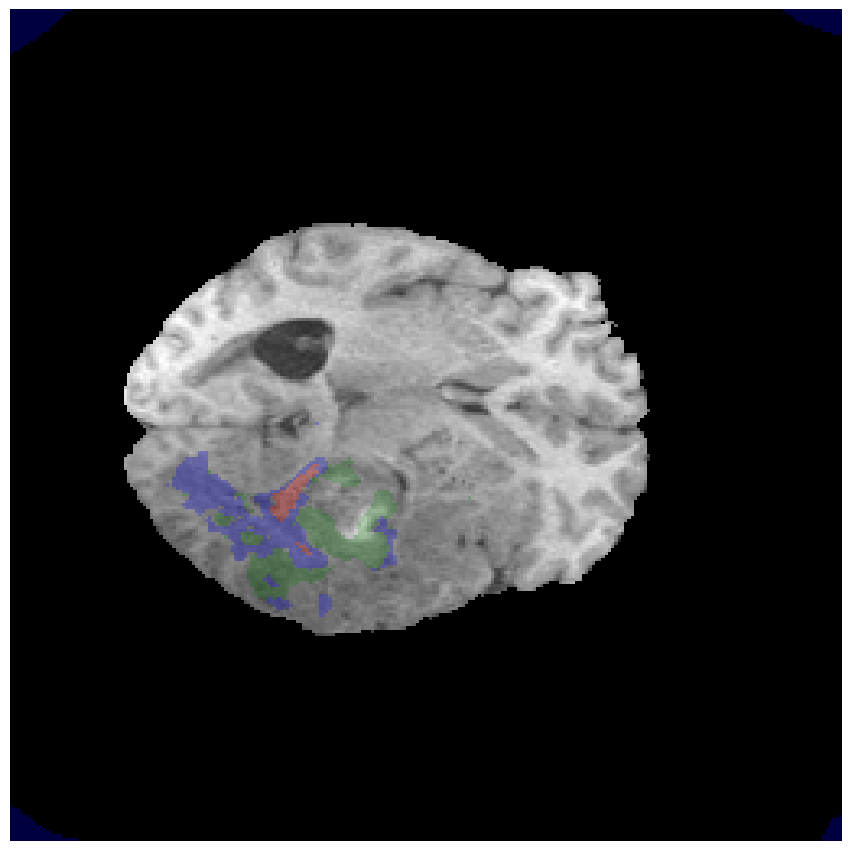} &
        \includegraphics[width=0.16\textwidth]{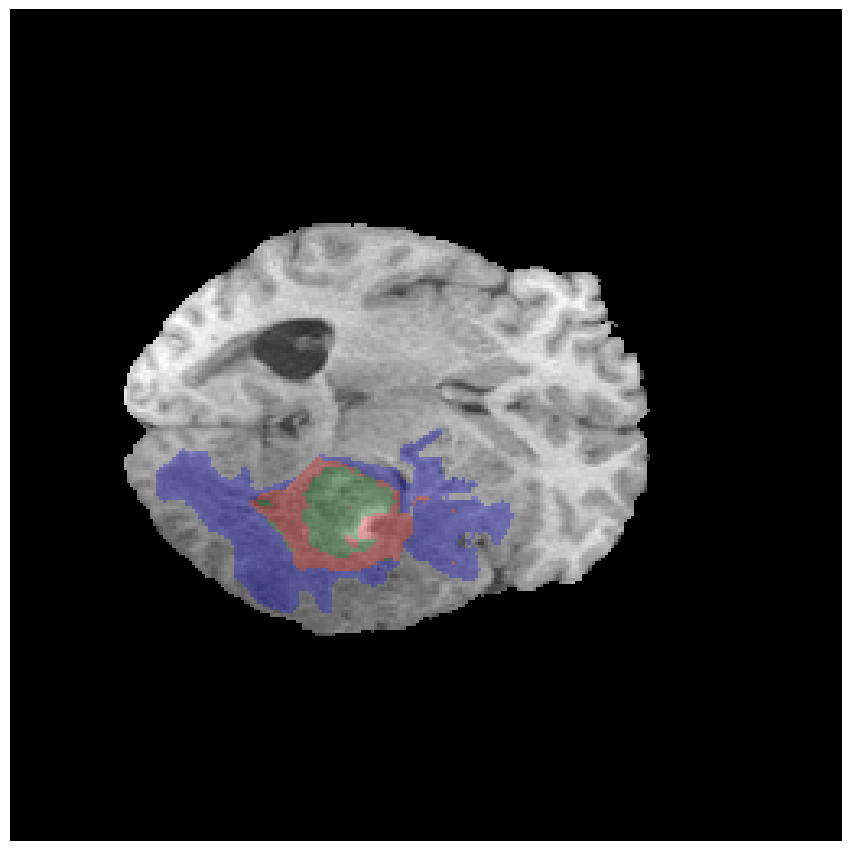} \\

    \end{tabular}
    \caption{Example U-Net predictions on an image in the BraTS 2020 test set. Classes are visualized as colored overlay where red is GD-enhancing
tumor, blue is peritumoral edema (ED) and green is necrotic and non-enhancing tumor core (NCR/NET). Each prediction is shown for four trainings using images from each generative model; with and without augmentation and with and without the original data. 
The two bottom rows present predictions from when training using synthetic images.
}
\label{fig:predictions_2020}
\end{figure*}

\begin{figure}
    \setlength{\tabcolsep}{0pt} 
    \begin{tabular}{cccccccc}
       
        & Annotation & None & StyleGAN 1 & StyleGAN 2 & StyleGAN 3 & Progressive GAN & Diffusion \\
        \rotatebox{90}{ Orig + Aug} &
        \includegraphics[width=0.16\textwidth]{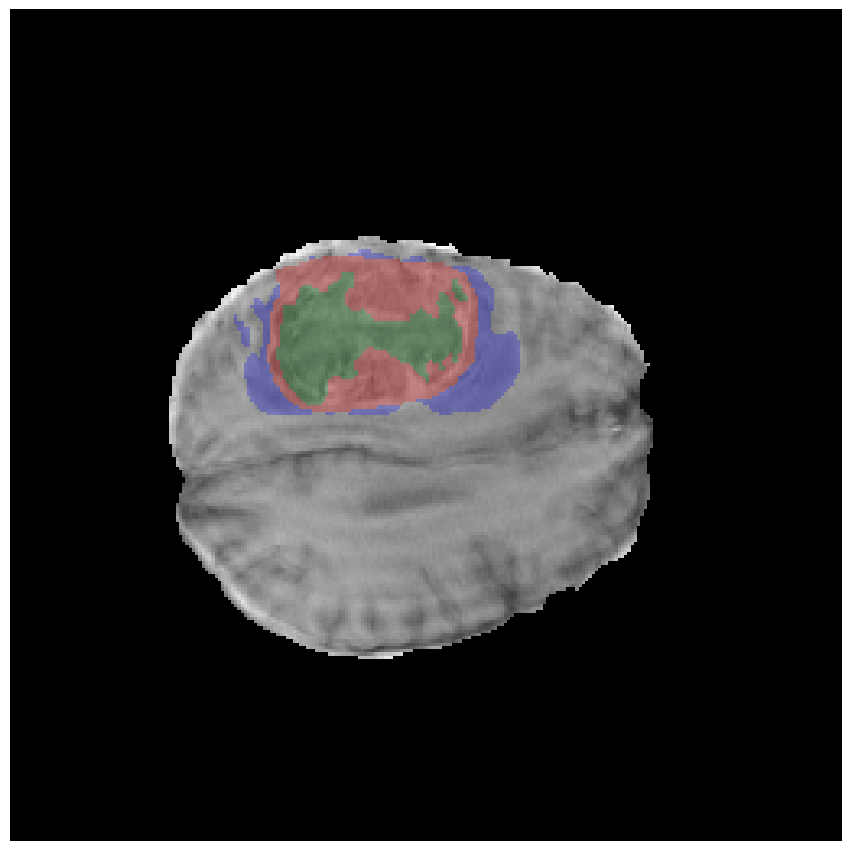} &
        \includegraphics[width=0.16\textwidth]{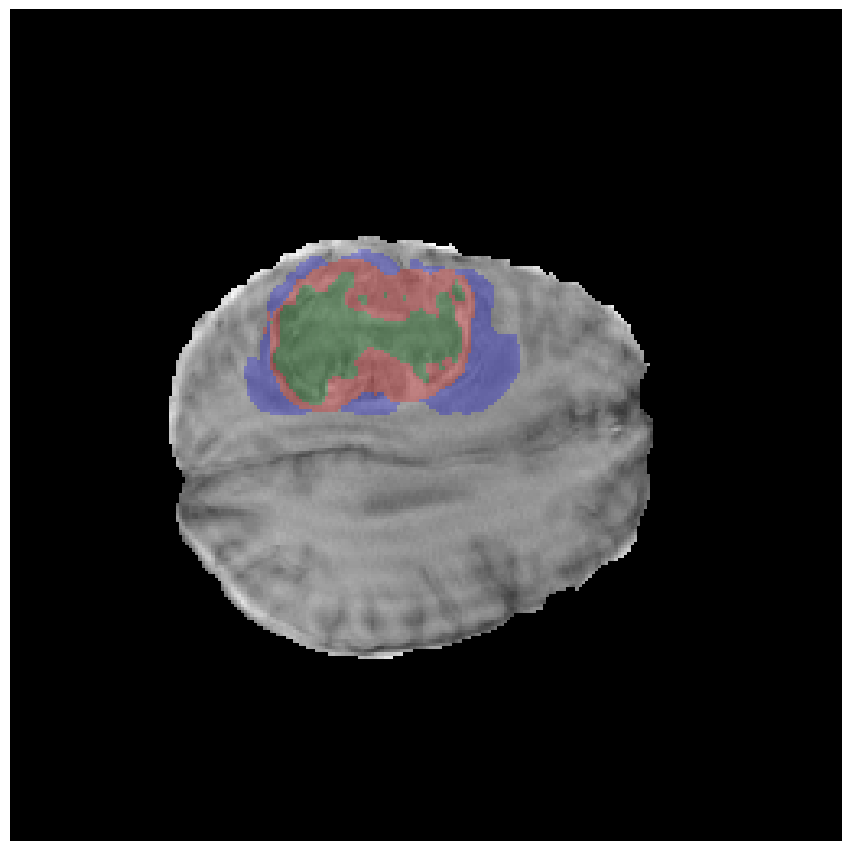} &
        \includegraphics[width=0.16\textwidth]{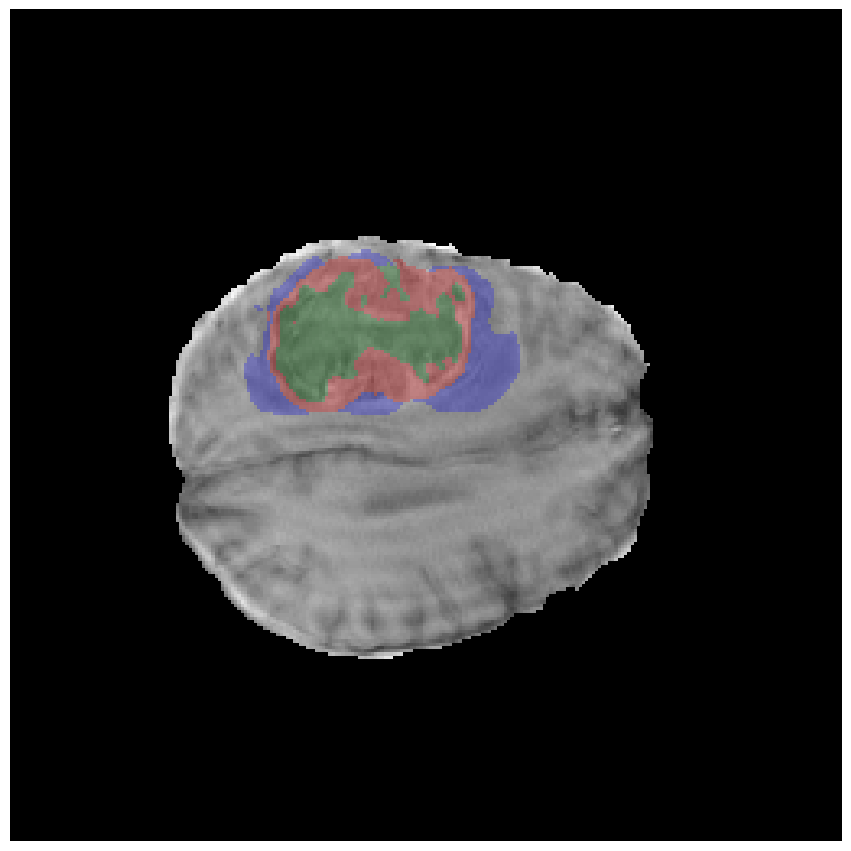} &
        \includegraphics[width=0.16\textwidth]{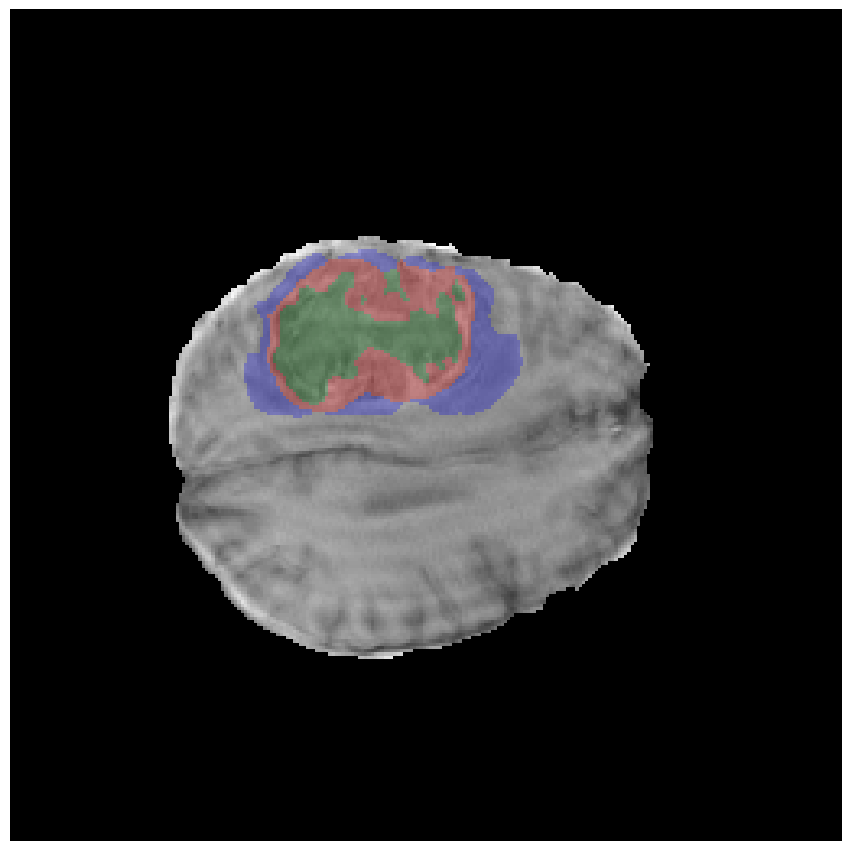} &
        \includegraphics[width=0.16\textwidth]{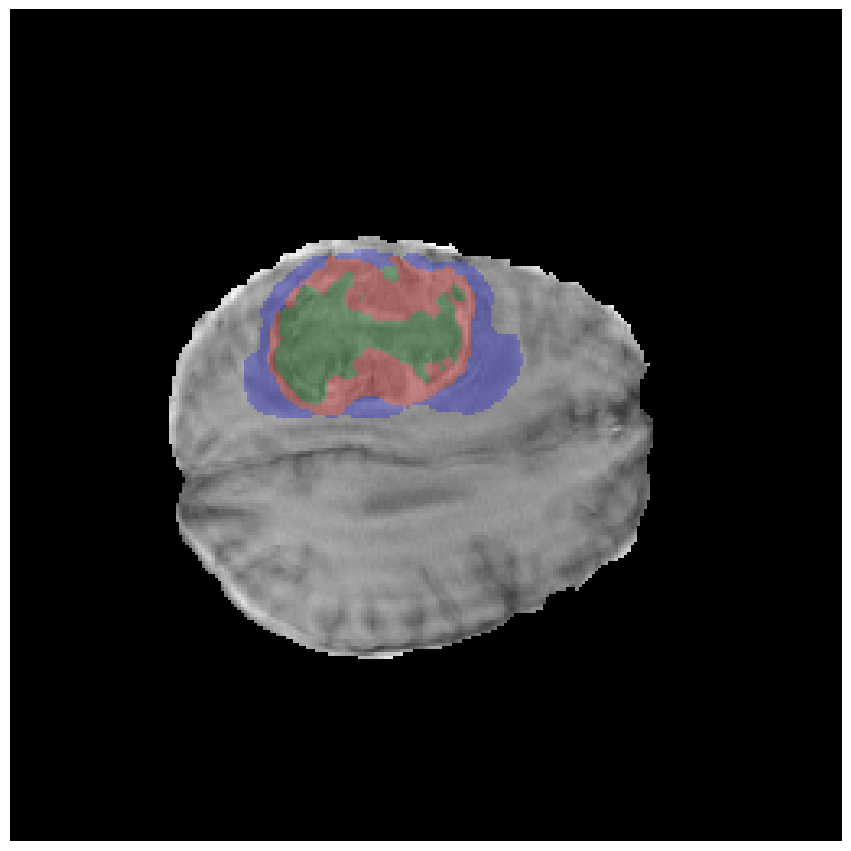} &
        \includegraphics[width=0.16\textwidth]{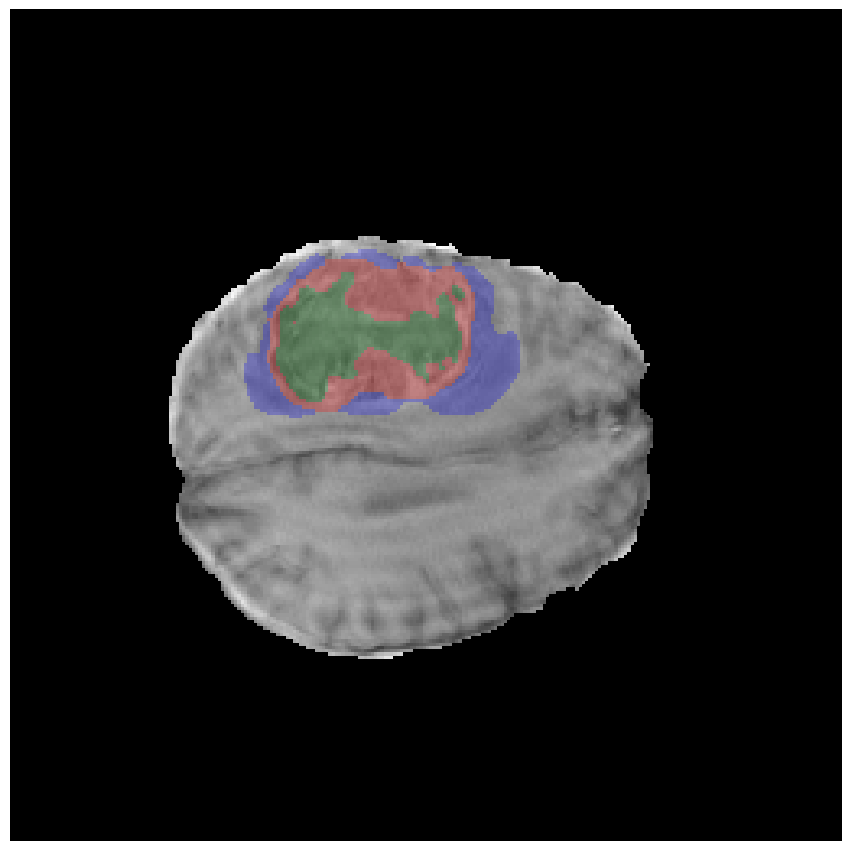} &
        \includegraphics[width=0.16\textwidth]{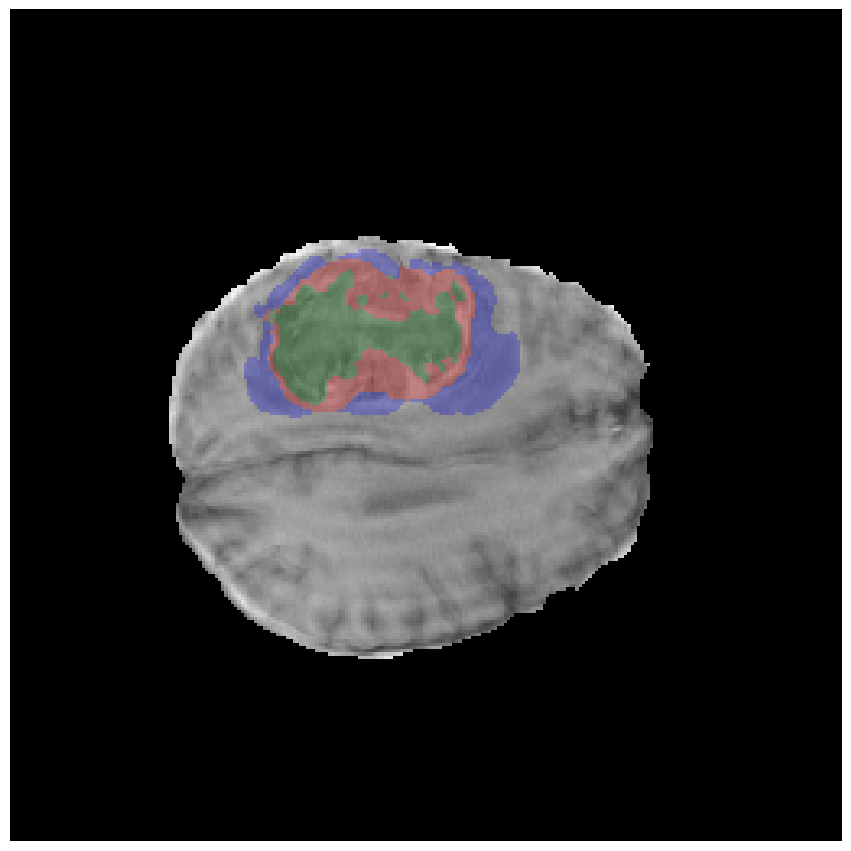} \\
        
        \rotatebox{90}{\hspace{.5cm} Orig} &
        \includegraphics[width=0.16\textwidth]{Figures/results/2021/Subject_0018_slice_101_T1/gt.png} &
        \includegraphics[width=0.16\textwidth]{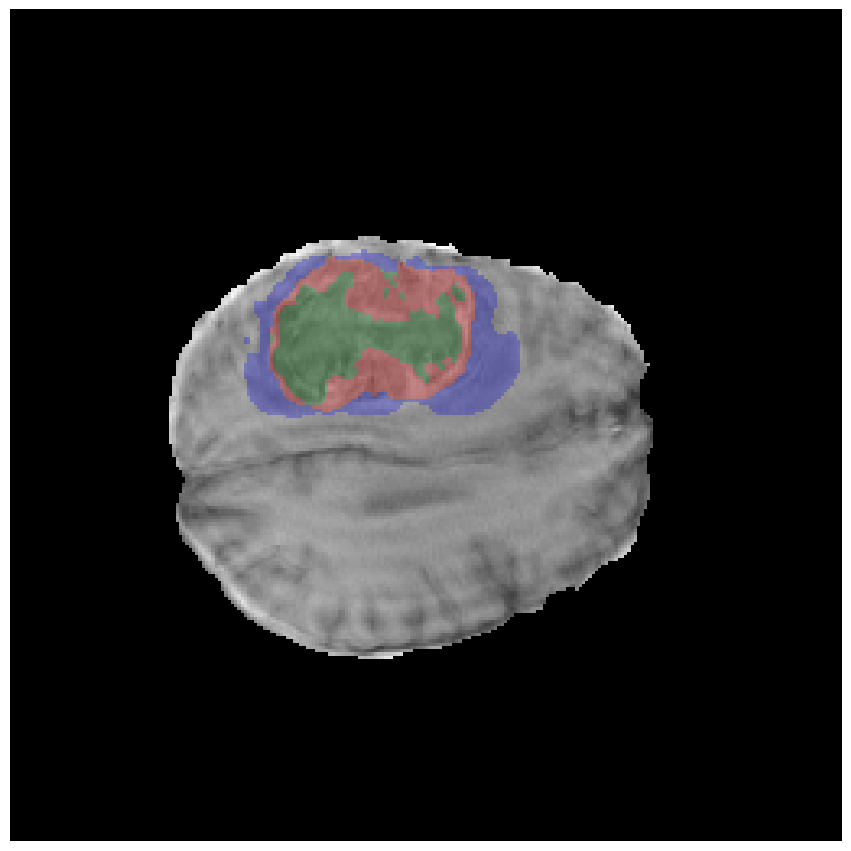} &
        \includegraphics[width=0.16\textwidth]{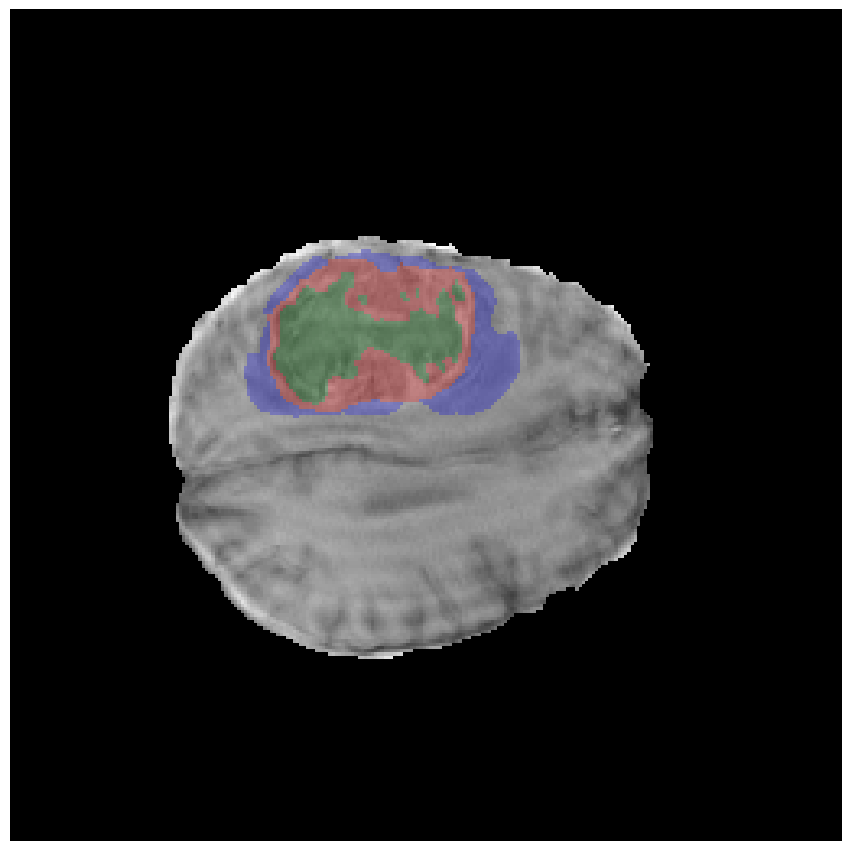} &
        \includegraphics[width=0.16\textwidth]{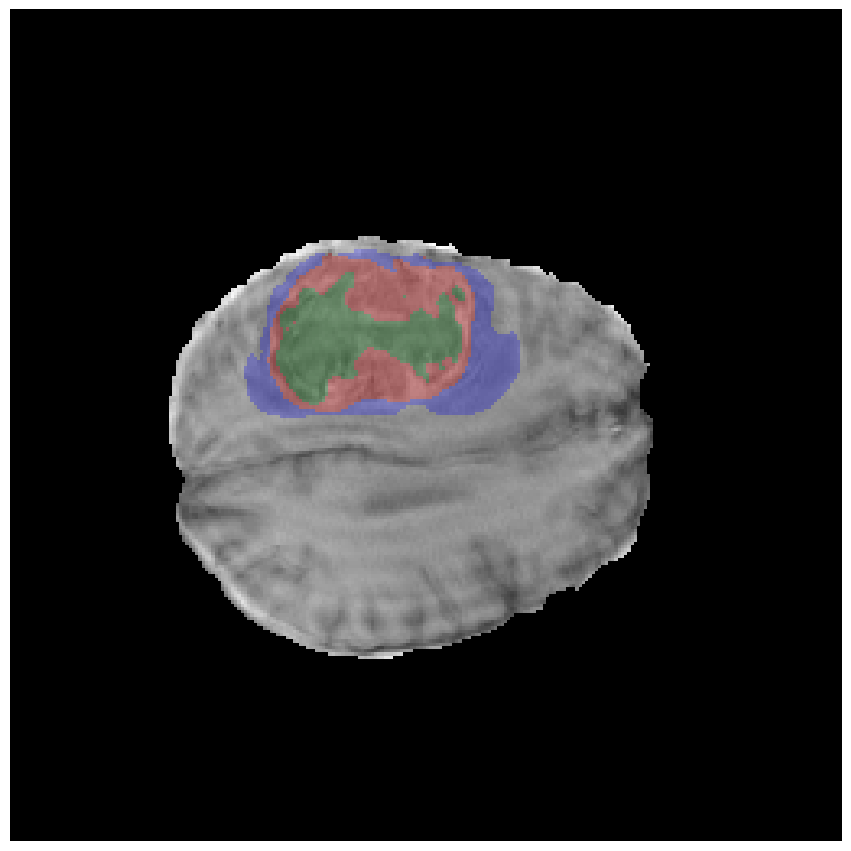} &
        \includegraphics[width=0.16\textwidth]{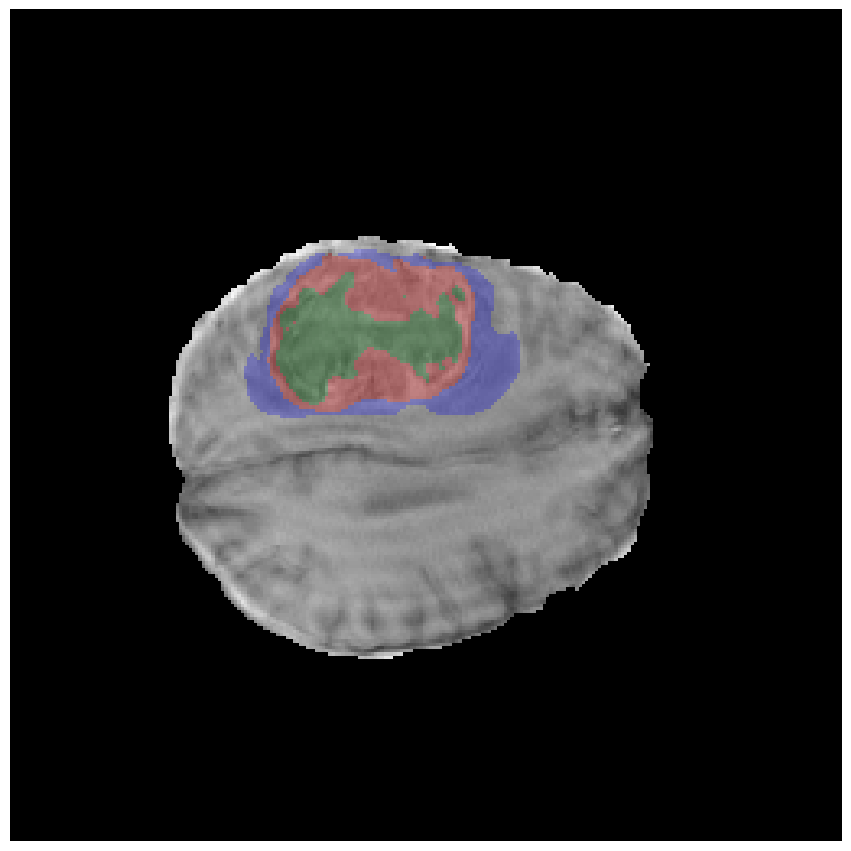} &
        \includegraphics[width=0.16\textwidth]{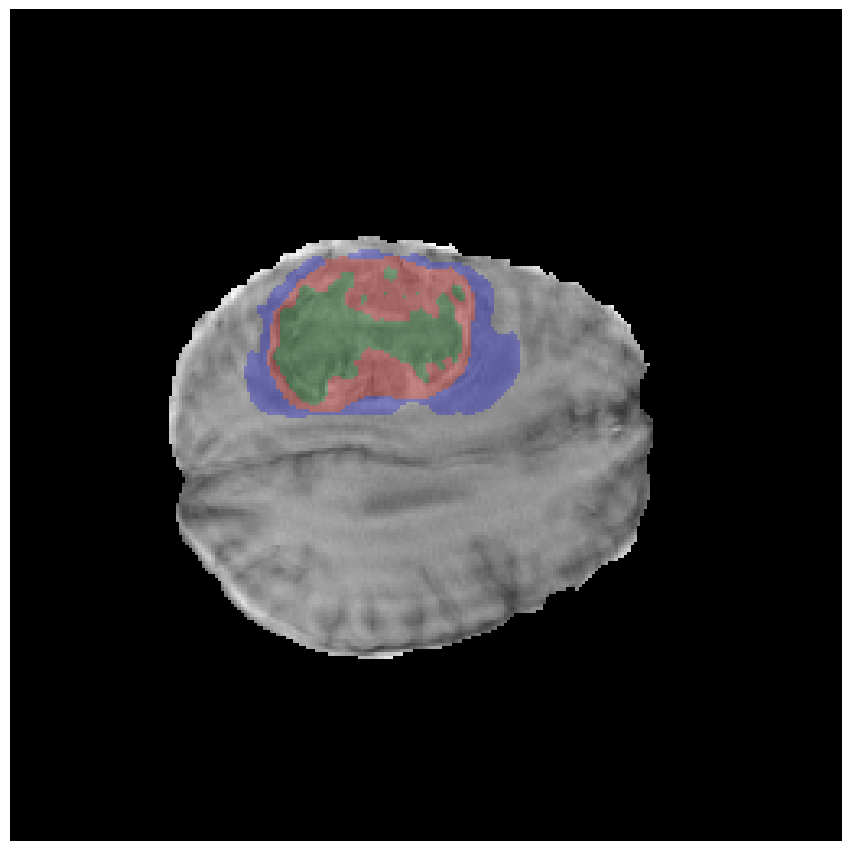} &
        \includegraphics[width=0.16\textwidth]{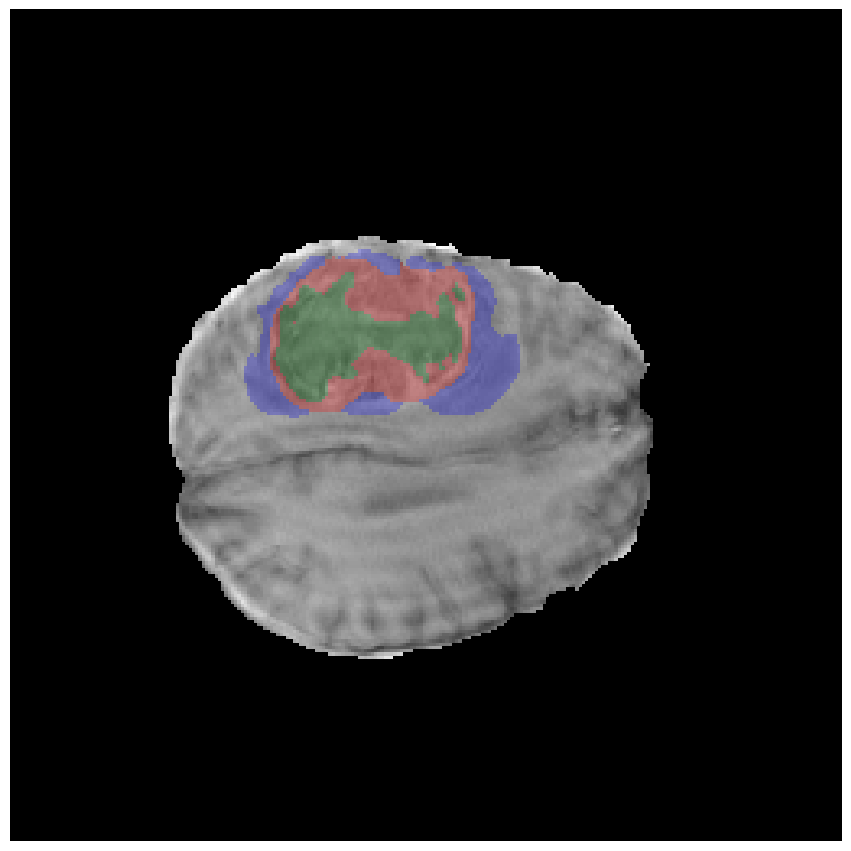} \\

        \rotatebox{90}{\hspace{.5cm} Aug} &
        \includegraphics[width=0.16\textwidth]{Figures/results/2021/Subject_0018_slice_101_T1/gt.png} &
         &
        \includegraphics[width=0.16\textwidth]{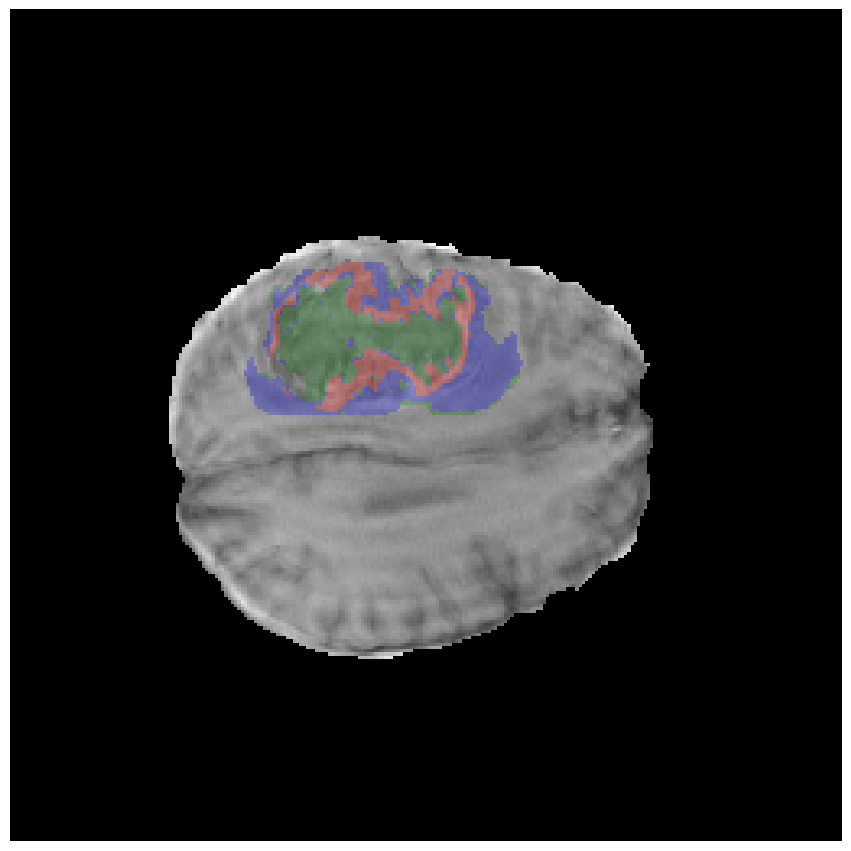} &
        \includegraphics[width=0.16\textwidth]{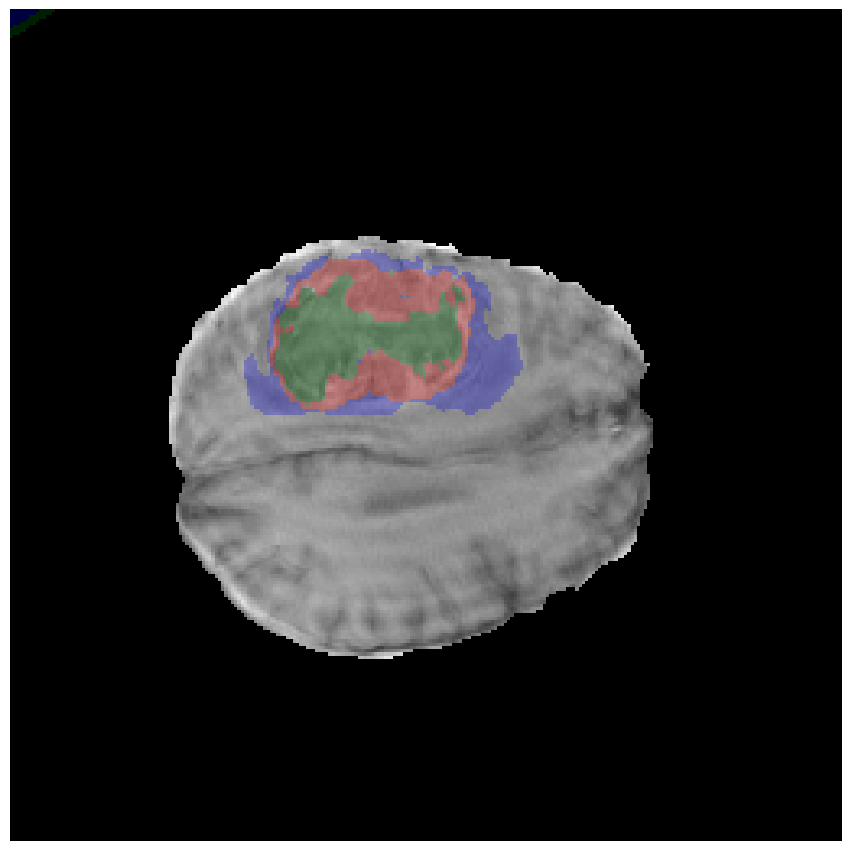} &
        \includegraphics[width=0.16\textwidth]{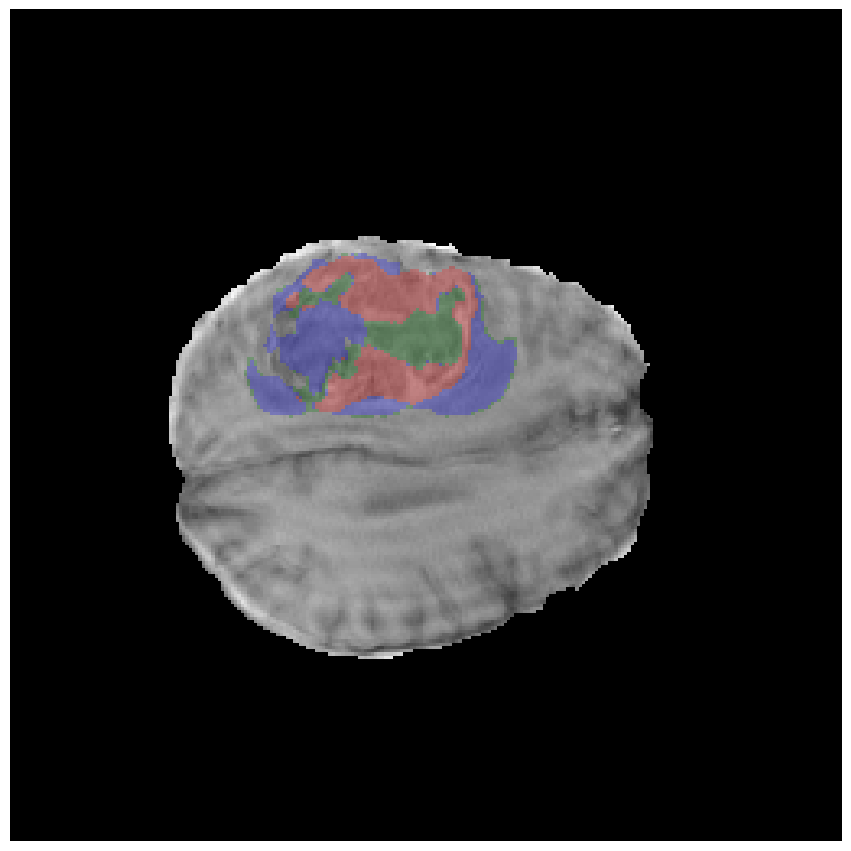} &
        \includegraphics[width=0.16\textwidth]{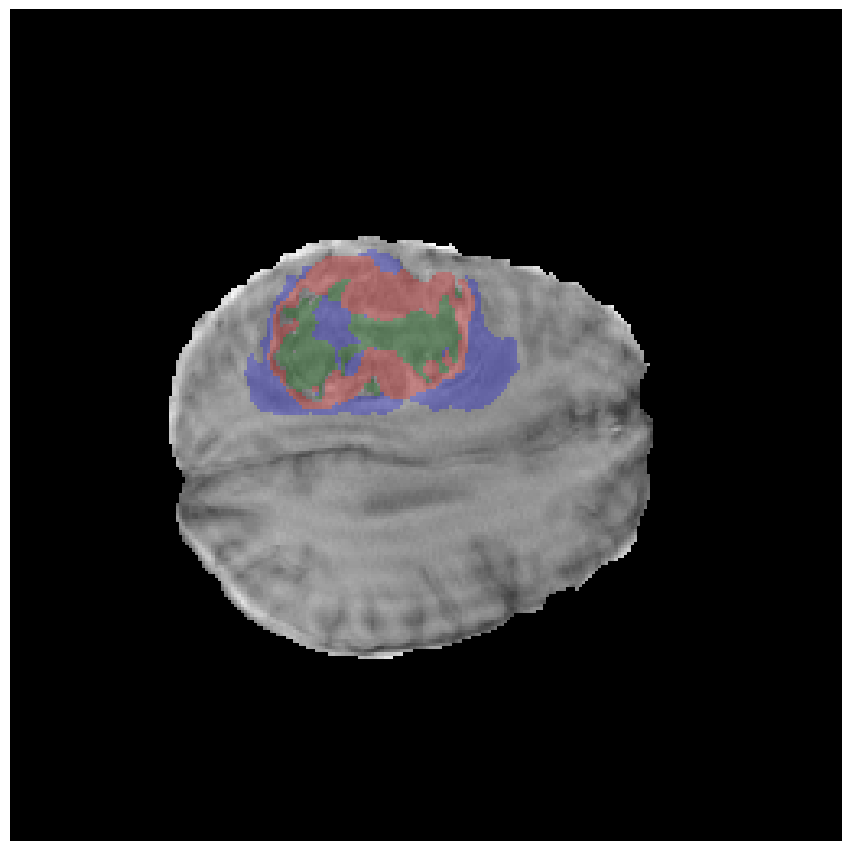} &
        \includegraphics[width=0.16\textwidth]{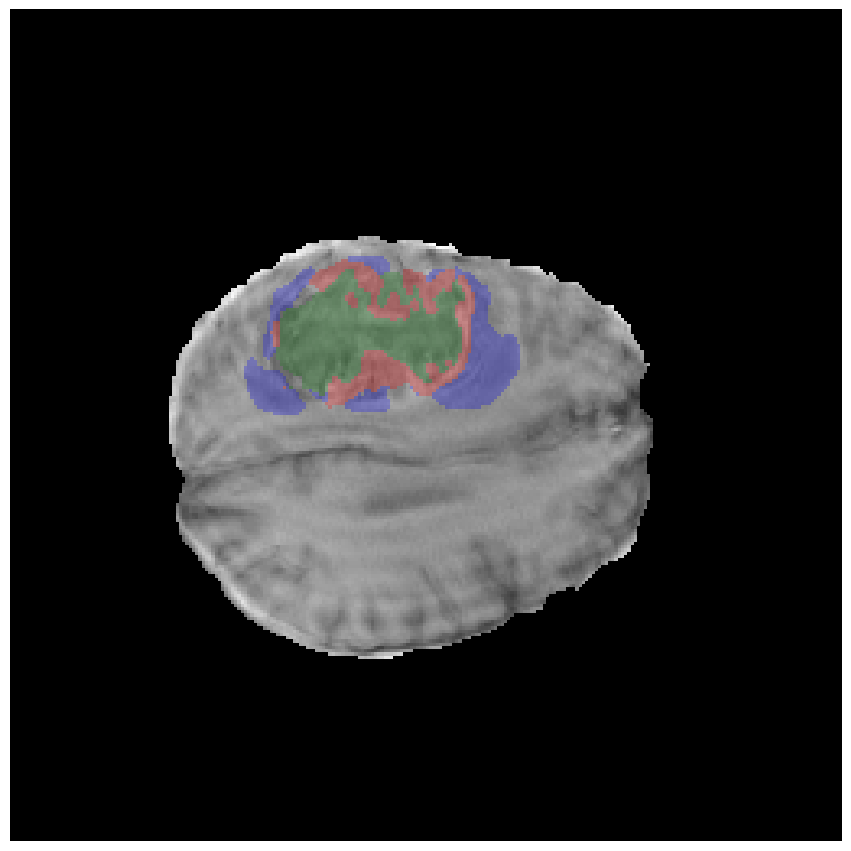} \\

        &
        \includegraphics[width=0.16\textwidth]{Figures/results/2021/Subject_0018_slice_101_T1/gt.png} &
         &
        \includegraphics[width=0.16\textwidth]{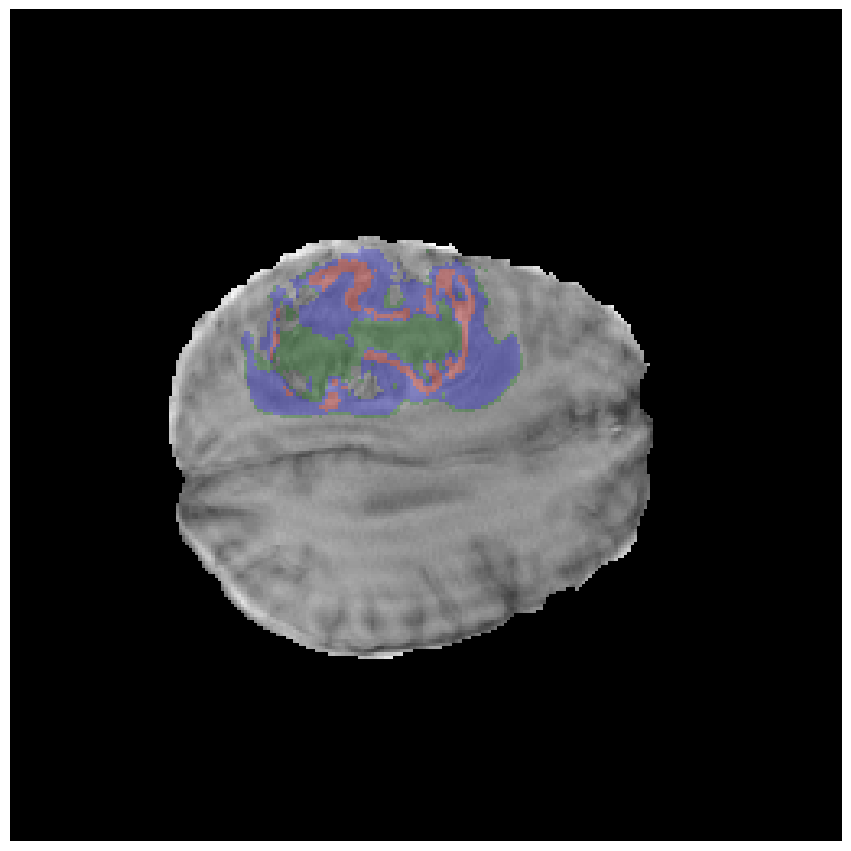} &
        \includegraphics[width=0.16\textwidth]{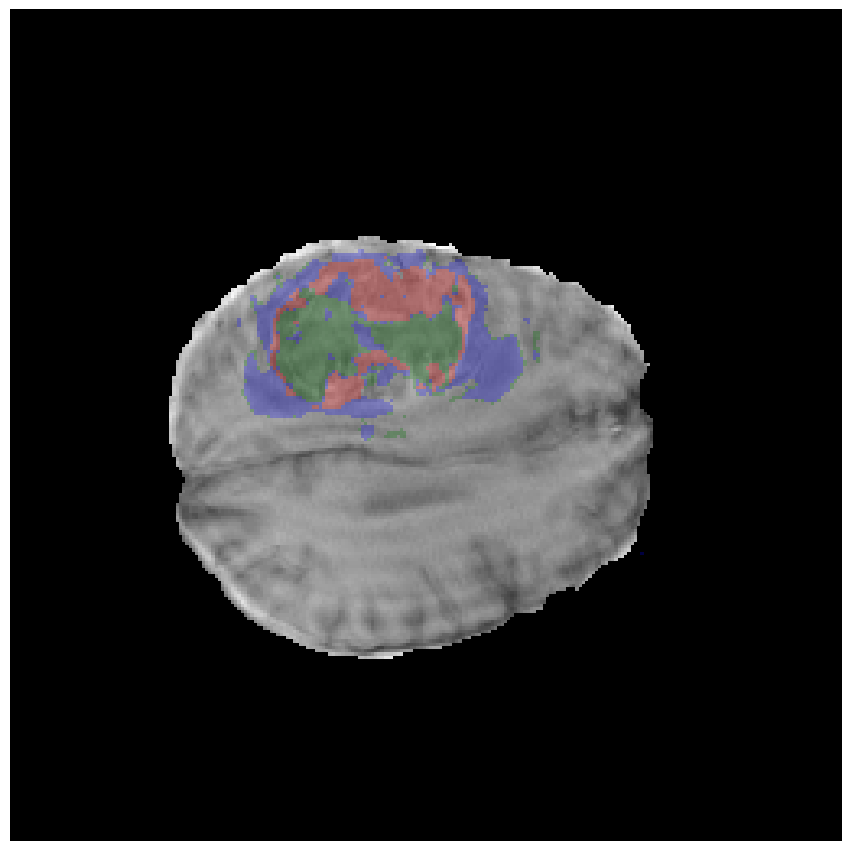} &
        \includegraphics[width=0.16\textwidth]{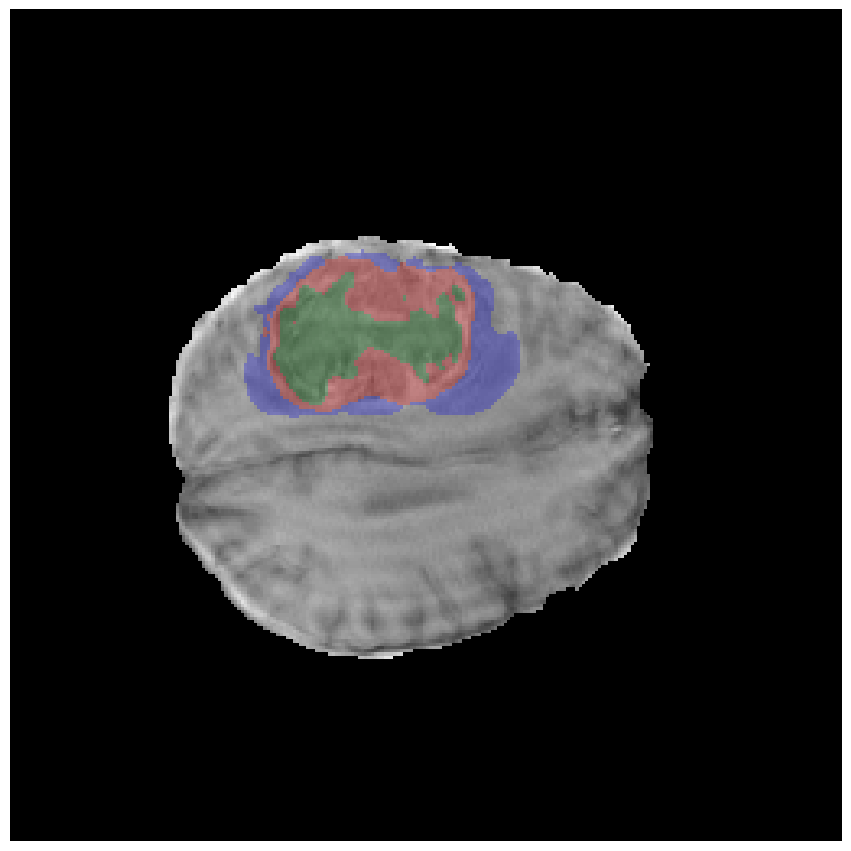} &
        \includegraphics[width=0.16\textwidth]{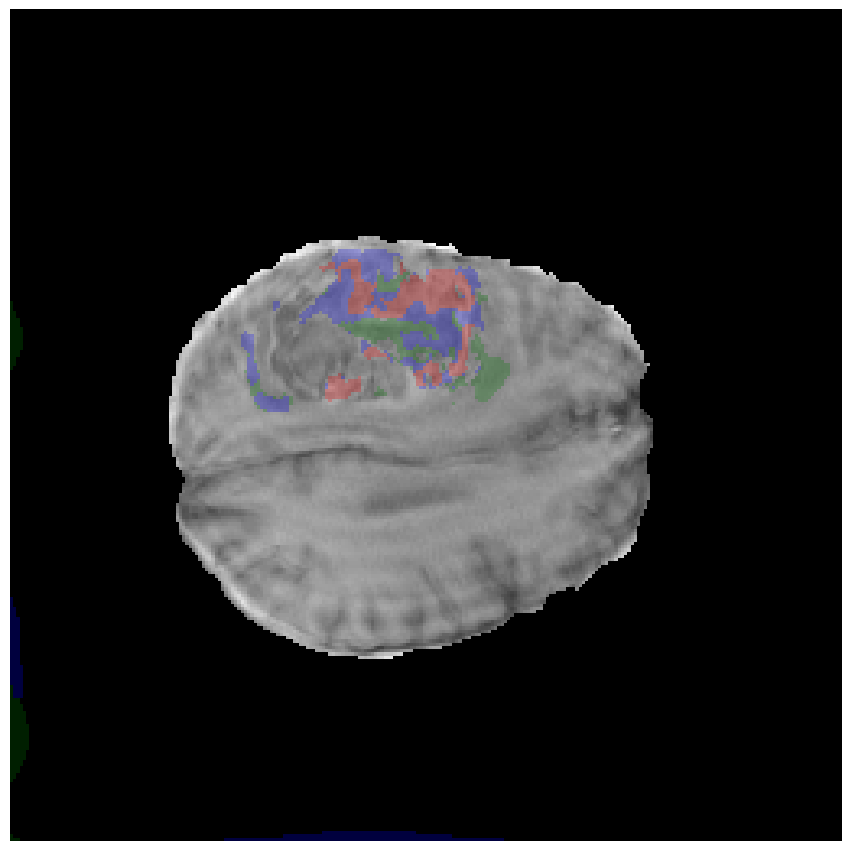} &
        \includegraphics[width=0.16\textwidth]{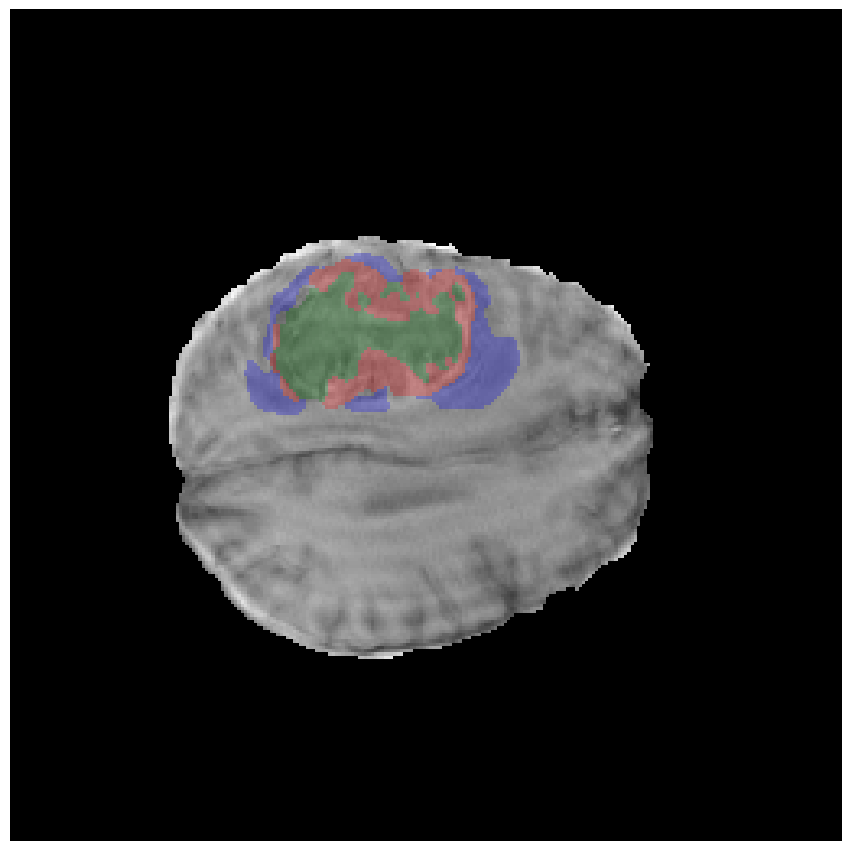} \\

    \end{tabular}
    \caption{Example U-Net predictions on an image in the BraTS 2021 test set. Classes are visualized as colored overlay where red is GD-enhancing
tumor, blue is peritumoral edema (ED) and green is necrotic and non-enhancing tumor core (NCR/NET). Each prediction is shown for four trainings using images from each generative model; with and without augmentation and with and without the original data. The two bottom rows present predictions from when training using synthetic images.}
\label{fig:predictions_2021}
\end{figure}
\begin{table*}  \scriptsize
    \centering
    \begin{tabular}{c|c|c|c|ccc|c} 
    
&        \textbf{Model} & \textbf{Orig} & \textbf{Aug} & \textbf{ET} & \textbf{ED} & \textbf{NCR/NET} & \textbf{Mean} \\ \hline
        
\multirow{22}{*}{\rotatebox{90}{\textbf{U-Net}}}
&None & \checkmark & \checkmark & $16.4 \pm 1.8$ & $33.8 \pm 3.4$ & $19.2 \pm 1.2$ & $23.1 \pm 1.9$ \\
&Progressive GAN & \checkmark & \checkmark & $15.9 \pm 1.6$ & $33.8 \pm 2.9$ & $21.5 \pm 1.9$ & $23.7 \pm 1.8$ \\
&StyleGAN 1 & \checkmark & \checkmark & $26.6 \pm 16.0$ & $42.6 \pm 11.1$ & $24.4 \pm 4.0$ & $31.2 \pm 9.9$ \\
&\textbf{StyleGAN 2} & \checkmark & \checkmark & $\bf{15.4 \pm 1.9}$ & $\bf{32.7 \pm 2.6}$ & $\bf{19.6 \pm 1.4}$ & $\bf{22.6 \pm 1.5}$ \\
&StyleGAN 3 & \checkmark & \checkmark & $31.3 \pm 22.7$ & $44.4 \pm 15.4$ & $33.7 \pm 19.2$ & $36.5 \pm 18.7$ \\
&Diffusion & \checkmark & \checkmark & $21.7 \pm 20.1$ & $32.3 \pm 8.4$ & $22.3 \pm 14.0$ & $25.4 \pm 14.1$ \\
&None & \checkmark & & $17.7 \pm 2.4$ & $35.9 \pm 4.5$ & $21.3 \pm 2.7$ & $24.9 \pm 2.8$ \\
&Progressive GAN & \checkmark & & $17.7 \pm 3.1$ & $35.2 \pm 4.2$ & $22.6 \pm 2.0$ & $25.1 \pm 2.5$ \\
&StyleGAN 1 & \checkmark & & $22.3 \pm 12.4$ & $39.1 \pm 9.1$ & $23.5 \pm 3.2$ & $28.3 \pm 7.8$ \\
&StyleGAN 2 & \checkmark & & $16.3 \pm 3.2$ & $34.2 \pm 2.8$ & $20.2 \pm 1.8$ & $23.6 \pm 2.0$ \\
&StyleGAN 3 & \checkmark & & $24.5 \pm 17.6$ & $39.2 \pm 12.3$ & $26.9 \pm 15.3$ & $30.2 \pm 14.7$ \\
&Diffusion & \checkmark & & $19.1 \pm 14.5$ & $32.9 \pm 6.5$ & $21.4 \pm 10.0$ & $24.5 \pm 10.1$ \\
&Progressive GAN & &  \checkmark & $35.2 \pm 6.7$ & $44.9 \pm 4.3$ & $34.0 \pm 4.3$ & $38.0 \pm 3.7$ \\
&StyleGAN 1 & &  \checkmark & $82.1 \pm 6.0$ & $90.1 \pm 40.5$ & $108.6 \pm 49.4$ & $93.6 \pm 29.6$ \\
&StyleGAN 2 & &  \checkmark & $33.4 \pm 7.9$ & $62.4 \pm 24.5$ & $36.4 \pm 11.4$ & $44.1 \pm 11.9$ \\
&StyleGAN 3 & &  \checkmark & $37.8 \pm 12.2$ & $49.6 \pm 7.5$ & $50.4 \pm 7.8$ & $45.9 \pm 8.9$ \\
&\textbf{Diffusion} & &  \checkmark & $\bf{15.2 \pm 7.1}$ & $\bf{32.1 \pm 3.9}$ & $\bf{26.3 \pm 10.6}$ & $\bf{24.6 \pm 5.8}$ \\
&Progressive GAN & & & $58.6 \pm 40.9$ & $94.6 \pm 58.3$ & $70.1 \pm 46.7$ & $74.4 \pm 44.9$ \\
&StyleGAN 1 & & & $103.7 \pm 28.7$ & $112.2 \pm 39.1$ & $131.2 \pm 44.5$ & $115.7 \pm 33.5$ \\
&StyleGAN 2 & & & $106.6 \pm 78.2$ & $116.8 \pm 61.7$ & $97.7 \pm 73.6$ & $107.0 \pm 68.0$ \\
&StyleGAN 3 & & & $48.8 \pm 14.6$ & $58.9 \pm 10.9$ & $64.0 \pm 17.7$ & $57.2 \pm 13.6$ \\
&Diffusion & & & $16.7 \pm 5.7$ & $34.9 \pm 5.4$ & $28.4 \pm 9.5$ & $26.6 \pm 5.2$ \\ \hline

\multirow{22}{*}{\rotatebox{90}{\textbf{Swin Transformer}}}
&None                & \checkmark    & \checkmark    & $16.1 \pm 1.2$    & $27.5 \pm 1.7$  & $19.3 \pm 1.4$   & $20.3 \pm 0.8$ \\
&Progressive GAN     & \checkmark    & \checkmark    & $25.9 \pm 2.1$    & $47.9 \pm 1.6$  & $47.6 \pm 2.3$   & $38.3 \pm 1.7$ \\
&StyleGAN 1          & \checkmark    & \checkmark    & $13.8 \pm 1.1$    & $26.1 \pm 0.9$  & $18.2 \pm 1.1$   & $18.7 \pm 1.0$ \\
&StyleGAN 2          & \checkmark    & \checkmark    & $14.5 \pm 1.1$    & $27.1 \pm 0.8$  & $19.0 \pm 1.3$   & $19.8 \pm 0.9$ \\
&StyleGAN 3          & \checkmark    & \checkmark    & $14.5 \pm 0.7$    & $25.2 \pm 0.7$  & $19.4 \pm 2.4$   & $19.3 \pm 1.0$ \\
&\textbf{Diffusion}           & \checkmark    & \checkmark    & $\textbf{14.5} \pm \textbf{0.9}$    & $\textbf{25.7} \pm \textbf{1.0}$  & $\textbf{19.4} \pm \textbf{1.1}$   & $\textbf{19.2} \pm \textbf{0.7}$ \\
&None                & \checkmark    &               & $16.8 \pm 0.6$    & $29.6 \pm 1.0$  & $20.0 \pm 1.0$   & $21.4 \pm 0.5$ \\
&Progressive GAN     & \checkmark    &               & $16.8 \pm 0.7$    & $30.4 \pm 1.3$  & $20.3 \pm 0.9$   & $21.6 \pm 0.8$ \\
&StyleGAN 1          & \checkmark    &               & $16.6 \pm 1.3$    & $28.3 \pm 1.1$  & $19.6 \pm 0.8$   & $20.6 \pm 0.6$ \\
&StyleGAN 2          & \checkmark    &               & $15.7 \pm 0.9$    & $28.8 \pm 1.7$  & $19.4 \pm 0.9$   & $20.4 \pm 0.9$ \\
&StyleGAN 3          & \checkmark    &               & $15.3 \pm 1.3$    & $26.8 \pm 1.4$  & $17.9 \pm 1.7$   & $19.4 \pm 1.2$ \\
&Diffusion           & \checkmark    &               & $16.0 \pm 0.6$    & $28.0 \pm 1.3$  & $19.7 \pm 0.7$   & $20.9 \pm 0.5$ \\
&Progressive GAN     &               & \checkmark    & $26.6 \pm 1.7$    & $48.5 \pm 1.0$  & $47.4 \pm 1.6$   & $38.4 \pm 0.9$ \\
&StyleGAN 1          &               & \checkmark    & $75.6 \pm 4.2$    & $72.3 \pm 2.1$  & $87.9 \pm 2.1$   & $74.8 \pm 3.1$ \\
&StyleGAN 2          &               & \checkmark    & $19.2 \pm 1.5$    & $37.8 \pm 2.0$  & $47.7 \pm 2.8$   & $33.5 \pm 1.7$ \\
&StyleGAN 3          &               & \checkmark    & $20.5 \pm 1.0$    & $38.8 \pm 1.5$  & $48.0 \pm 1.6$   & $34.0 \pm 1.4$ \\
&\textbf{Diffusion}           &               & \checkmark    & $\textbf{14.7} \pm \textbf{0.8}$    & $\textbf{25.2} \pm \textbf{1.0}$  & $\textbf{20.0} \pm \textbf{1.2}$   & $\textbf{19.5} \pm \textbf{0.7}$ \\
&Progressive GAN     &               &               & $38.7 \pm 1.9$    & $63.3 \pm 1.1$  & $69.9 \pm 1.9$   & $54.6 \pm 0.4$ \\
&StyleGAN 1          &               &               & $74.6 \pm 2.8$    & $66.5 \pm 2.7$  & $82.7 \pm 2.7$   & $70.0 \pm 2.3$ \\
&StyleGAN 2          &               &               & $47.4 \pm 3.1$    & $64.0 \pm 0.5$  & $79.1 \pm 0.9$   & $60.1 \pm 1.2$ \\
&StyleGAN 3          &               &               & $23.4 \pm 2.7$    & $39.6 \pm 2.2$  & $51.6 \pm 1.7$   & $36.7 \pm 1.9$ \\
&Diffusion           &               &               & $16.2 \pm 0.9$    & $27.5 \pm 1.3$  & $20.3 \pm 1.3$   & $20.8 \pm 0.9$ \\
    \end{tabular}
    \vspace{0.2cm}
    \caption{Results on the test dataset (56 subjects) when training the generative models with BraTS 2020 (313 training subjects). Mean and standard deviation of Hausdorff distance across the dataset are calculated for the labels; GD-enhancing tumor (ET), peritumoral edema (ED), and necrotic and non-enhancing tumor core (NCR/NET). All results are presented as mean $\pm$ standard deviation of 10 trainings. The column Orig marks if the original dataset has been added to the training set and the column Aug marks if augmentation was used during training of the segmentation network. Training with and without augmentation was performed with the same number of total images. The top row shows baseline results using only real images. The bottom 10 rows show results when training with only synthetic images, with and without augmentation. }
    \label{tab:res_ens_brats2020_haus}
\end{table*}

\begin{table*}  \scriptsize
    \centering
    \begin{tabular}{c|c|c|c|ccc|c} 
    
&        \textbf{Model} & \textbf{Orig} & \textbf{Aug} & \textbf{ET} & \textbf{ED} & \textbf{NCR/NET} & \textbf{Mean} \\ \hline
        \multirow{22}{*}{\rotatebox{90}{\textbf{U-Net}}} 
&None & \checkmark & \checkmark & $12.3 \pm 2.0$ & $22.2 \pm 2.1$ & $12.1 \pm 1.7$ & $15.5 \pm 1.6$ \\
&Progressive GAN & \checkmark & \checkmark & $30.5 \pm 49.1$ & $30.6 \pm 14.9$ & $12.9 \pm 2.7$ & $24.7 \pm 21.2$ \\
&\textbf{StyleGAN 1} & \checkmark & \checkmark & $\bf{13.3 \pm 0.4}$ & $\bf{23.5 \pm 3.4}$ & $\bf{12.0 \pm 1.2}$ & $\bf{16.2 \pm 1.4}$ \\
&StyleGAN 2 & \checkmark & \checkmark & $25.7 \pm 36.2$ & $37.4 \pm 47.0$ & $25.4 \pm 42.7$ & $29.5 \pm 41.9$ \\
&StyleGAN 3 & \checkmark & \checkmark & $13.8 \pm 4.4$ & $24.8 \pm 7.9$ & $12.8 \pm 1.0$ & $17.1 \pm 4.2$ \\
&Diffusion & \checkmark & \checkmark & $16.2 \pm 3.3$ & $26.6 \pm 4.5$ & $12.4 \pm 1.2$ & $18.4 \pm 1.6$ \\
&None & \checkmark & & $12.8 \pm 1.7$ & $22.5 \pm 2.2$ & $11.9 \pm 1.3$ & $15.7 \pm 1.3$ \\
&Progressive GAN & \checkmark & & $21.6 \pm 35.9$ & $28.4 \pm 10.9$ & $12.3 \pm 2.1$ & $20.8 \pm 15.5$ \\
&StyleGAN 1 & \checkmark & & $18.9 \pm 21.6$ & $29.1 \pm 24.2$ & $18.8 \pm 30.0$ & $22.3 \pm 25.2$ \\
&StyleGAN 2 & \checkmark & & $19.2 \pm 26.5$ & $30.4 \pm 34.0$ & $18.5 \pm 30.9$ & $22.7 \pm 30.4$ \\
&StyleGAN 3 & \checkmark & & $14.7 \pm 5.4$ & $26.4 \pm 7.8$ & $12.6 \pm 1.6$ & $17.9 \pm 4.6$ \\
&Diffusion & \checkmark & & $14.9 \pm 3.1$ & $25.0 \pm 3.9$ & $12.3 \pm 0.9$ & $17.4 \pm 1.8$ \\
&Progressive GAN & &  \checkmark & $31.2 \pm 5.3$ & $48.0 \pm 4.4$ & $23.9 \pm 1.6$ & $34.3 \pm 3.0$ \\
&StyleGAN 1 & &  \checkmark & $52.8 \pm 14.2$ & $76.7 \pm 31.1$ & $99.8 \pm 21.3$ & $76.4 \pm 15.9$ \\
&StyleGAN 2 & &  \checkmark & $78.5 \pm 53.3$ & $124.6 \pm 46.2$ & $115.8 \pm 57.6$ & $106.3 \pm 41.5$ \\
&StyleGAN 3 & &  \checkmark & $30.8 \pm 7.3$ & $50.5 \pm 3.1$ & $63.6 \pm 3.5$ & $48.3 \pm 3.2$ \\
&Diffusion & &  \checkmark & $28.8 \pm 8.8$ & $41.9 \pm 4.6$ & $15.1 \pm 4.1$ & $28.6 \pm 3.9$ \\
&Progressive GAN & & & $86.9 \pm 64.8$ & $112.8 \pm 65.1$ & $108.6 \pm 84.9$ & $102.7 \pm 69.6$ \\
&StyleGAN 1 & & & $108.6 \pm 57.3$ & $124.6 \pm 53.1$ & $142.8 \pm 46.9$ & $125.3 \pm 50.7$ \\
&StyleGAN 2 & & & $106.0 \pm 48.1$ & $142.4 \pm 40.6$ & $151.3 \pm 55.7$ & $133.2 \pm 41.2$ \\
&StyleGAN 3 & & & $31.0 \pm 5.7$ & $52.2 \pm 3.6$ & $67.0 \pm 4.8$ & $50.0 \pm 3.6$ \\
&\textbf{Diffusion} & & & $\bf{31.7 \pm 14.7}$ & $\bf{38.0 \pm 5.9}$ & $\bf{15.5 \pm 3.7}$ & $\bf{28.4 \pm 5.4}$ \\ \hline
\multirow{22}{*}{\rotatebox{90}{\textbf{Swin Transformer}}}
&None                & \checkmark    & \checkmark    & $11.3 \pm 0.4$    & $17.2 \pm 0.7$  & $11.4 \pm 0.8$   & $11.5 \pm 0.3$ \\
&Progressive GAN     & \checkmark    & \checkmark    & $11.5 \pm 0.9$    & $18.6 \pm 0.9$  & $11.2 \pm 0.9$   & $11.7 \pm 0.6$ \\
&\textbf{StyleGAN 1 }         & \checkmark    & \checkmark    & $\textbf{10.4} \pm \textbf{0.8}$    & $\textbf{18.1} \pm \textbf{0.9}$  & $\textbf{10.9} \pm \textbf{0.5}$   & $\textbf{11.1} \pm \textbf{0.3}$ \\
&StyleGAN 2          & \checkmark    & \checkmark    & $11.1 \pm 0.7$    & $17.2 \pm 1.1$  & $11.1 \pm 0.5$   & $11.2 \pm 0.5$ \\
&StyleGAN 3          & \checkmark    & \checkmark    & $11.2 \pm 0.9$    & $17.5 \pm 1.1$  & $11.1 \pm 0.5$   & $11.3 \pm 0.6$ \\
&Diffusion           & \checkmark    & \checkmark    & $11.0 \pm 0.5$    & $16.9 \pm 0.6$  & $11.5 \pm 0.6$   & $11.2 \pm 0.5$ \\
&None                & \checkmark    &               & $11.4 \pm 0.9$    & $18.1 \pm 1.0$  & $11.4 \pm 0.9$   & $11.6 \pm 0.8$ \\
&Progressive GAN     & \checkmark    &               & $11.3 \pm 0.6$    & $18.9 \pm 0.8$  & $11.1 \pm 0.6$   & $11.7 \pm 0.7$ \\
&StyleGAN 1          & \checkmark    &               & $11.1 \pm 0.7$    & $18.1 \pm 1.2$  & $11.1 \pm 0.3$   & $11.2 \pm 0.6$ \\
&StyleGAN 2          & \checkmark    &               & $11.6 \pm 0.9$    & $17.3 \pm 1.3$  & $11.2 \pm 0.5$   & $11.2 \pm 0.9$ \\
&StyleGAN 3          & \checkmark    &               & $11.2 \pm 0.3$    & $18.1 \pm 0.9$  & $11.2 \pm 0.3$   & $11.8 \pm 0.4$ \\
&Diffusion           & \checkmark    &               & $11.5 \pm 0.7$    & $17.3 \pm 1.1$  & $11.5 \pm 0.7$   & $11.6 \pm 0.3$ \\
&Progressive GAN     &               & \checkmark    & $26.3 \pm 1.8$    & $49.9 \pm 1.3$  & $55.0 \pm 1.4$   & $40.2 \pm 1.0$ \\
&StyleGAN 1          &               & \checkmark    & $18.1 \pm 1.5$    & $41.1 \pm 1.9$  & $62.5 \pm 1.4$   & $37.1 \pm 1.10$ \\
&StyleGAN 2          &               & \checkmark    & $14.0 \pm 0.8$    & $31.6 \pm 1.7$  & $42.5 \pm 2.1$   & $25.7 \pm 1.0$ \\
&StyleGAN 3          &               & \checkmark    & $16.0 \pm 1.3$    & $38.4 \pm 2.6$  & $44.5 \pm 1.2$   & $29.2 \pm 1.3$ \\
&\textbf{Diffusion}           &               & \checkmark    & $\textbf{12.7} \pm \textbf{0.7}$    & $\textbf{17.7} \pm \textbf{0.6}$  & $\textbf{14.1} \pm \textbf{0.5}$   & $\textbf{12.7} \pm \textbf{0.6}$ \\
&Progressive GAN     &               &               & $49.2 \pm 1.9$    & $63.4 \pm 0.7$  & $72.8 \pm 0.8$   & $58.2 \pm 0.8$ \\
&StyleGAN 1          &               &               & $24.3 \pm 1.8$    & $58.9 \pm 3.1$  & $73.9 \pm 0.7$   & $49.1 \pm 1.2$ \\
&StyleGAN 2          &               &               & $19.3 \pm 1.2$    & $47.0 \pm 1.6$  & $58.9 \pm 1.5$   & $38.4 \pm 1.1$ \\
&StyleGAN 3          &               &               & $16.8 \pm 1.6$    & $43.3 \pm 1.9$  & $49.6 \pm 2.2$   & $32.5 \pm 1.4$ \\
&Diffusion           &               &               & $13.7 \pm 0.9$    & $19.2 \pm 0.8$  & $13.3 \pm 0.7$   & $13.0 \pm 0.8$ \\

    \end{tabular}
    \vspace{0.2cm}
    \caption{Results on the test dataset (56 subjects) when training the generative models with BraTS 2021 (1195 training subjects). Mean and standard deviation of Hausdorff distance across the dataset are calculated for the labels; GD-enhancing tumor (ET), peritumoral edema (ED), and necrotic and non-enhancing tumor core (NCR/NET). All results are presented as mean $\pm$ standard deviation of 10 trainings. The column Orig marks if the original dataset has been added to the training set and the column Aug marks if augmentation was used during training of the segmentation network. Training with and without augmentation was performed with the same number of total images. The top row shows baseline results using only real images. The bottom 10 rows show results when training with only synthetic images, with and without augmentation. }
    \label{tab:res_ens_brats2021_haus}
\end{table*}

\end{document}